\documentclass[journal]{IEEEtran}

\usepackage{cite}
\usepackage{amsmath,amssymb,amsfonts}
\usepackage{algorithmic}
\usepackage{graphicx}
\usepackage{textcomp}
\usepackage{xcolor}
\usepackage{multicol}
\usepackage{caption}
\usepackage{subcaption}
\usepackage{balance}
\newcommand{\C}{{\cal C}}
\newcommand{\D}{{\cal D}}

\newcommand{\X}{{\cal X}}
\newcommand{\qed}{\vspace{-5pt}\begin{flushright}$\blacksquare$\end{flushright}\vspace{-10pt}}
\allowdisplaybreaks

\newtheorem{theorem}{Theorem}

\newtheorem{proposition}[theorem]{Proposition}

\begin{document}

\title{Optimal Binary Signaling for a Two Sensor Gaussian MAC Network\\

\thanks{The authors are with the Department of Mathematics and Statistics, Queen's University,
Kingston, Ontario, Canada (\{16ls53,takahara,fa\}@queensu.ca)).
This work was supported in part by the Natural Sciences and Engineering Research Council (NSERC) of Canada.}
}

\author{Luca Sardellitti, Glen Takahara, Fady Alajaji}

\maketitle

\begin{abstract}
We consider a two sensor distributed detection system transmitting a binary non-uniform source over a Gaussian multiple access channel (MAC). We model the network via binary sensors whose outputs are generated by binary symmetric channels of different noise levels. We prove an optimal one dimensional constellation design under individual sensor power constraints which minimizes the error probability of detecting the source. Three distinct cases arise for this optimization based on the parameters in the problem setup. In the most notable case (Case~III), the optimal signaling design is to not necessarily use all of the power allocated to the more noisy sensor (with less correlation to the source). We compare the error performance of the optimal one dimensional constellation to orthogonal signaling. The results show that the optimal one dimensional constellation achieves lower error probability than using orthogonal channels.
\end{abstract}

\begin{IEEEkeywords}
Distributed detection, wireless sensor networks, multiple access channel, constellation design, binary signaling, power allocation, source-channel signaling, error probability.
\end{IEEEkeywords}

\section{Introduction}
Wireless sensor networks are widely  used for monitoring the state of real world phenomena. This includes both the estimation of a real valued parameter (such as temperature or rain fall measurements) and the detection of an event occurring (such as the occurrence of forest fires or a security breach). In this paper, we focus on the hypothesis testing problem described by distributed detection of an event occurring.

When working with generalized distributed detection problems, the error probability of the system cannot generally be expressed analytically. As a result, previous work on distributed detection typically uses related or proxy metrics for system error analysis. For example, \cite{errorExponents} uses error exponents to evaluate the performance of various detection schemes, \cite{powerAllocOrthoChannels} uses the J-divergence (i.e., the Jeffreys-divergence~\cite{jeffreys1946invariant}),  while \cite{thetaOptimize} and \cite{deflectCoeff} use the deflection coefficient as the metric for optimization. 

Previous work in this area employs a variety of signaling structures for the sensors. For example, \cite{powerAllocOrthoChannels} uses orthogonal channels for each sensor, while \cite{deflectCoeff} uses a single MAC for the entire network. Works such as \cite{deflectCoeff, ferrari14, detectionAlgo, jiang05} fix a signaling design and analyze detection schemes at the fusion center. Other works optimize the sensors' signaling techniques under certain constraints. For example, \cite{powerAllocOrthoChannels} optimizes power allocation for a network that uses orthogonal signaling and on-off keying for each sensor, under total and average power constraints. Alternatively, \cite{thetaOptimize} assumes fixed power at each sensor, and analyzes the optimal rotation angle to send the signals.

Throughout the above mentioned works, there is not much emphasis put on generalized constellation design for distributed detection problems. We aim to solve the source-channel signaling problem of finding an optimized constellation design to minimize error probability under a given source and channel model. This is similar in principle to works such as \cite{weng18, lin19, minDistOptimize, jointConstMimo, powerAndModulation, 2DConstDesign, M-aryConstOpt, pamOptm, mAryOptForm}, where general constellation design is optimized for a chosen criterion. In~\cite{weng18}, the optimal joint binary constellation design for two correlated sources was derived. In~\cite{lin19} and \cite{minDistOptimize}, the authors used a minimum inter-constellation distance criterion for optimizing constellations for multiple sources. In contrast, \cite{2DConstDesign,M-aryConstOpt, pamOptm, mAryOptForm} optimized $M$-ary constellations for a single source.
In this paper, we adapt these ideas to optimize signaling for a distributed detection system.

We simplify the distributed detection problem to a two sensor network so that an analytical optimization of the error probability can be performed. We model hypothesis testing for an event of interest occurring as a non-uniformly distributed binary source, and the sensor noises are modelled as passing the source through independent memoryless binary symmetric channels, introducing sensor errors. Finally, the sensors choose binary constellations to send their signals over a Gaussian MAC to the fusion center, which then performs the maximum-a-posteriori (MAP) detection rule to recover the source. With this setup, we analytically derive an optimal one dimensional constellation design to minimize the system error probability under individual power constraints for each sensor. The proof is split into three cases, based on the parameters describing the source distribution and sensors' noise. The constellation design is optimized in each of these cases using indirect analysis of the decision boundaries, followed by algebraic and derivative analysis using an upper bound on the error probability. In addition, we show using numerical and simulation results that the derived optimal one dimensional constellation achieves lower error probability than using orthogonal channels. Our most notable result is that in certain cases (which is dominant when the source is nearly uniformly distributed), the noisier sensor should use some but not all of its allocated power. 

The rest of the paper is structured as follows. Section~\ref{ModelSect} describes the mathematical model of the sensor network, including the fusion center detection rule. Sections~\ref{resultSummary} and~\ref{proofSect} give a summary and a detailed proof of the main optimization results, respectively. Section~\ref{numericalSect} presents simulations and numerical examples which reinforce and illustrate the theoretically established results. Finally, Section~\ref{Conclusions} draws conclusions and suggests future research directions.
\section{System Model}\label{ModelSect}
\subsection{Source and Sensors}
Let $X$ be a binary event that is to be observed by a sensor network. Without loss of generality, it is assumed that the source is distributed such that $p_1 \triangleq \text{Pr}(X=1) \leq 0.5$. We also define $p_0 \triangleq \text{Pr}(X = 0) = 1 - p_1$. There are two sensors, $X_1$ and $X_2$ observing the source $X$, which are modelled as passing $X$ through two memoryless binary symmetric channels. This is expressed as $X_s = X \oplus Z_s,\; s=1,2$, where $\oplus$ denotes addition modulo-2, with $Z_1$ and $Z_2$ being independent Bernoulli noise processes 
with means (or channel crossover probabilities) $\epsilon_1$ and $\epsilon_2$, respectively. It is also assumed that $X$ is independent from $(Z_1,Z_2)$. Without loss of generality, Sensor 1 is assumed to have stronger correlation to the original source $X$ than Sensor~2: $0<\epsilon_1 \leq \epsilon_2<0.5$. The sensors, unable to communicate with each other, encode their data independently using binary constellations. The constellations for the sensors are represented as follows: $\C_s = \{c_{0,s}, c_{1,s}\}$, $s\in \{1,2\}$, where for $i\in \{0,1\}$,  $c_{i,s} \in \mathbb{R}$ denotes the constellation point for sensor $s$ assigned to $X_s = i$. Let $S_1\in \C_1$, $S_2 \in \C_2$ be the random variables associated to each sensor's chosen constellation point. Also let $P_1^{\text{max}}$ and $P_2^{\text{max}}$ be the power constraints of sensors, i.e., $E[S_i^2] \leq P_i^{\text{max}},\; i\in\{1,2\}$. In this setup, each sensor has its own power allotment, as opposed to having a common power constraint on the entire network.
\subsection{Channel Model}
The sensors' signals are sent through a Gaussian MAC. The received signal $R$ is described by the relation $R = S_1 + S_2 + Z$, where $Z$ is a Gaussian noise variable with zero mean and variance $\frac{N_0}{2}$. For convenience, we also define $\sigma$ as the standard deviation of the noise, given by $\sigma = \sqrt{\frac{N_0}{2}}$. It is assumed that $Z$ is independent of the sensor signals $S_1$ and $S_2$. The overall signal $S_1+S_2$ sent over the channel can be represented as a point in the combined constellation of $\C_1$ and $\C_2$, given by $\C = \{c_1 + c_2 \:|\: c_1 \in \C_1,\: c_2 \in \C_2\}$.
\subsection{Maximum-a-Posteriori Detection}
The event $X$ is reconstructed at the fusion center using (optimal) MAP detection. For a Gaussian MAC received signal $r$, the detected bit is determined as follows:
\begin{align} 
    \hat{x}(r) &= \underset{i\in\{0,1\}}{\arg\max}\; \text{Pr}(X = i\;|\;R = r) \nonumber \\
    &= \underset{i\in\{0,1\}}{\arg\max}\; \text{Pr}(X = i)f_R(r\;|\;X=i) \nonumber \\
   & = \underset{i\in\{0,1\}}{\arg\max}\;p_i\hspace{-10pt}\sum_{(l,m)\in\{0,1\}^2}\hspace{-10pt}p_{lm|i}f_R(r|S_1+S_2=a_{lm})\nonumber \\
    & = \underset{i\in\{0,1\}}{\arg\max}\;p_i\hspace{-10pt}\sum_{(l,m)\in\{0,1\}^2}\hspace{-10pt}p_{lm|i}f_Z(r-a_{lm}),
   \label{MAPDecodeExpression}
\end{align}
where $f_R$ and $f_Z$ are the probability density functions (pdfs) of the received signal $R$ and channel noise variable $Z$, respectively, $p_{lm|i} \triangleq\text{Pr}(X_1 = l, X_2 = m | X = i)$, and $a_{lm} \in \C$ denotes the superimposed constellation symbol associated with $X_1 = l$ and $X_2 = m$. In the case of a tie, we choose to detect a 0. This is an arbitrary decision because the probability of a tie is always zero since the noise is a continuous random variable. The conditional probabilities $p_{lm|i}$ can be expressed as follows in terms of the sensor crossover probabilities:
\begin{equation} \label{conditionalProbs}
\begin{aligned}
    p_{11|0} &= p_{00|1} = \epsilon_1\epsilon_2, \;\;
    p_{00|0} = p_{11|1} = (1-\epsilon_1)(1-\epsilon_2), \\    
    p_{01|0} &= p_{10|1} = (1-\epsilon_1)\epsilon_2,\;\;
    p_{10|0} = p_{01|1} = \epsilon_1(1-\epsilon_2).
\end{aligned}
\end{equation}
The real line is partitioned into two decision regions, $\D_0$ and $\D_1=\D_0^c$, where $\D_i = \{r\in\mathbb{R}\;|\;\hat{x}(r) = i\}$, $i=0,1$.

\section{Summary of Main Results} \label{resultSummary}
We herein summarize our main results. Specifically, 
for fixed parameters $p_1$, $\epsilon_1$, 
$\epsilon_2$, $N_0$, $P_1^{\text{max}}$ and $P_2^{\text{max}}$, we show that the optimal constellation design for $\C_1$ and $\C_2$ which minimizes error probability are expressed as $\C_i = \{c_{0,i}, c_{1,i}\} = \left\{-\sqrt{\frac{p_1}{p_0}}P_i^*,\; \sqrt{\frac{p_0}{p_1}}P_i^*\right\}$, for $i\in \{1,2\}$, where the optimal power allocations $P_i^*$ are separated into three cases. The conditions for each case are given in Table~\ref{caseCharConditions} and the optimization results are summarized in Table~\ref{optimizationSummaryTable}. 
\begin{table}[htbp]
\caption{Case Characterization Conditions}
\label{caseCharConditions}
\begin{center}
\begin{tabular}{||c||c||}
\hline\hline
Case&Condition \\\hline
\hline
\rule{0pt}{2.6ex}
I & $0 \leq p_1 \leq \frac{\epsilon_1\epsilon_2}{1-\epsilon_1-\epsilon_2+2\epsilon_1\epsilon_2}$  \\ \hline
\rule{0pt}{2.6ex}
II & $\frac{\epsilon_1\epsilon_2}{1-\epsilon_1-\epsilon_2+2\epsilon_1\epsilon_2} < p_1 \leq \frac{\epsilon_1-\epsilon_1\epsilon_2}{\epsilon_1+\epsilon_2-2\epsilon_1\epsilon_2}$ \\ \hline
\rule{0pt}{2.6ex}
III & $\frac{\epsilon_1-\epsilon_1\epsilon_2}{\epsilon_1+\epsilon_2-2\epsilon_1\epsilon_2} < p_1 \leq 0.5$ \\ \hline\hline
\end{tabular}
\end{center}
\end{table}
\begin{table}[htbp]
\caption{Optimal Power Allocation Results}
\label{optimizationSummaryTable}
\begin{center}
\begin{tabular}{||c||c|c|c|c||}
\hline\hline
Case&$P_1^*$&$P_2^*$&$P_e^*$&$\lim_{N_0 \rightarrow 0}P_e^*$ \\\hline
\hline
\rule{0pt}{2.6ex}
I & 0  & 0 & $p_1$ & $p_1$\\\hline
\rule{0pt}{2.6ex}
II & $\sqrt{P_1^{\text{max}}}$ & $\sqrt{P_2^{\text{max}}}$ & see \eqref{PeCase2} & see \eqref{case2HighSNR}\\ \hline
\rule{0pt}{2.6ex}
III & $\sqrt{P_1^{\text{max}}}$ & $\min(\sqrt{P_2^{\text{max}}}, \Tilde{P}_2)$ & see \eqref{PeCase3} & $\epsilon_1$\\
\hline\hline
\end{tabular}
\end{center}
\end{table}

In Table~\ref{optimizationSummaryTable}, we used the quantities
\begin{equation} \label{nonUniformP2TildeMax}
    \Tilde{P}_2 \triangleq  \frac{N_0p_1p_0}{2\sqrt{P_1^{\text{max}}}}\ln\frac{(1 - \epsilon_1 - \epsilon_2)^2 - \Lambda}{(\epsilon_2 - \epsilon_1)^2 - \Lambda},
\end{equation}
with
\begin{equation*}
    \Lambda \triangleq \frac{(p_0-p_1)^2}{p_0p_1}(1-\epsilon_1)(1-\epsilon_2)\epsilon_1\epsilon_2.
\end{equation*}
Note that the expression for $\Tilde{P}_2$ is always real valued when the conditions of Case~III are met. Also note that because of the assumption $0<\epsilon_1\leq\epsilon_2<0.5$, the conditions for the three cases are numerically consistent, i.e., we have that
\begin{equation*}
    0 < \frac{\epsilon_1\epsilon_2}{1-\epsilon_1-\epsilon_2+2\epsilon_1\epsilon_2} < \frac{\epsilon_1-\epsilon_1\epsilon_2}{\epsilon_1+\epsilon_2-2\epsilon_1\epsilon_2} \leq 0.5,
\end{equation*}
where the last inequality holds with equality if and only if $\epsilon_1 = \epsilon_2$. These results will be further analyzed and discussed in later sections of this paper. The most interesting and counter-intuitive result is that in Case~III, the optimal power allocation is not to necessarily use all of the available power for Sensor 2. The remainder of this paper is dedicated to prove these results and  illustrate them via numerical examples and simulations.
\section{Proof of Main Results} \label{proofSect}
First, we use Theorem~\ref{asymmetricParameter} to show that the constellation design optimization problem can be restricted to a set of asymmetric constellations, parameterized by each sensor's power allocation. Then we analyze the boundary points between the decision regions $\D_0$ and $\D_1$ (decision boundaries) using this optimal asymmetric design. The characterization of these decision boundaries splits the problem into the three cases given in Table~\ref{caseCharConditions}. For Case~I, Proposition~\ref{case1nosol} shows the trivial nature of the decision boundaries, and thus also the optimization in this case. For Case~II, Propositions~\ref{xBoundCase2} and \ref{PropCase2P1XInfBound} derive bounds on the decision boundaries which are used in Theorem~\ref{Case2DecreaseP1P2} to show that using all allocated power for both sensors is optimal. Finally, in Case~III, Propositions~\ref{xBoundsP2Case3} and \ref{Case3P2XInfBoundProp} establish properties of the decision boundaries which are used in Theorem~\ref{Case3P2ErrorMinimization} to give the globally optimal power allocation for Sensor 2. Combining this result with Theorems~\ref{Case3P2decreasingError} and \ref{Case3DecreaseP1}, which show that the error probability decreases in both sensor powers until reaching the global minimum, yields the overall optimal power allocation under the given power constraints.
\begin{theorem} \label{asymmetricParameter}
For any combination of binary constellations $\C = \C_1 + \C_2$, there exists a constellation pattern $\C^* = \C_1^* + \C_2^*$ which has equal error probability, equal or better power consumption, with the composition $\C_i^* = \{-\sqrt{\frac{p_1}{p_0}}P_i,\; \sqrt{\frac{p_0}{p_1}}P_i\}$, for some $P_i \in \mathbb{R}$, $i \in \{1,2\}$.
\end{theorem}

\begin{IEEEproof}
In a Gaussian MAC using MAP detection, the error probability is the same for constellations that are translations of each other. Hence, constellations with the same distances between constellation points will have the same error performance. The distances between the points in the joint constellation $\C$ are determined by the distances between the points in the individual constellations $\C_i = \{c_{0,i}, c_{1,i}\}$, $\: i\in \{1,2\}$. Let the constellation distance, $d_i$ be defined as follows:
\begin{equation} \label{constellationDistance}
   d_i \triangleq c_{1,i} - c_{0,i}, \: i\in\{1,2\}.
\end{equation}
We will minimize the average power consumption for each sensor while maintaining constellation distance $d_i$. This is computed below substituting in the constraint from \eqref{constellationDistance}:
\begin{equation*}
    P_i^2 = E[S_i^2] = p_0c_{0,i}^2 + p_1c_{1,i}^2 = c_{1,i}^2-2p_0d_ic_{1,i}+p_0d_i^2.
\end{equation*}
This is a simple quadratic function of $c_{1,i}$, which is minimized at $c_{1,i} = p_0d_i$, $c_{0,i} = -p_1d_i$. Substituting this back into the expression of $P_i^2$, we see that the minimum power has the form ${P_i^*}^2 = p_1p_0d_i^2$. Finally, rearranging for $d_i = \frac{1}{\sqrt{p_1p_0}}P_i^*$ shows that the minimum power constellation has the form:
\begin{equation*}
    c_{0,i} = -\sqrt{\frac{p_1}{p_0}}P_i, \qquad c_{1,i} = \sqrt{\frac{p_0}{p_1}}P_i, \qquad i \in \{1,2\}.
\end{equation*}
\end{IEEEproof}
Using the result of Theorem~\ref{asymmetricParameter}, we can restrict the optimization search to constellations which take the following asymmetric form. $\C_i = \{c_{0,i}, c_{1,i}\} = \Big\{-\sqrt{\frac{p_1}{p_0}}P_i,\; \sqrt{\frac{p_0}{p_1}}P_i\Big\}$, where $P_i \in [0,\sqrt{P_i^{\text{max}}}]$ for $i\in \{1,2\}$. To simplify notation, we define the following two symbols which represent these optimal asymmetric parameters:
\begin{equation} \label{alphaBetaDefs}
    \alpha \triangleq \sqrt{\frac{p_0}{p_1}}, \qquad \beta \triangleq \sqrt{\frac{p_1}{p_0}}.
\end{equation}
The relationship $\alpha + \beta = \frac{1}{\sqrt{p_0p_1}}$ will be used often in the remaining analysis. The problem has now been reduced to finding the optimal power allocations, $P_1$ and $P_2$.
\subsection{Decision Boundaries}
To analyze the error probabilities, we first must understand the behaviour of the decision regions $\D_0$ and $\D_1$. To characterize these regions, we take the difference between the two terms in \eqref{MAPDecodeExpression} and manipulate the expressions to give
\begin{align*}
    &\quad\quad \text{Pr}(X = 1\;|\;R = r) - \text{Pr}(X = 0\;|\;R = r)\\ &= \sum_{(l,m)\in\{0,1\}^2}(p_1p_{lm|1}-p_0p_{lm|0})f_Z(r-a_{lm})\\
    &=  \frac{1}{\sigma\sqrt{2\pi}}\sum_{(l,m)\in\{0,1\}^2}(p_1p_{lm|1}-p_0p_{lm|0})e^{\frac{-(r-a_{lm})^2}{N_0}}.
\end{align*}
We are interested in the sign of this expression, so we simplify using the following forms of the restricted constellation points: 
\begin{align*}
    a_{11} = \alpha(P_1 + P_2), &\qquad a_{01} = -\beta P_1 + \alpha P_2, \\
    a_{10} = \alpha P_1 -\beta P_2, &\qquad a_{00} = -\beta( P_1 + P_2),
\end{align*}
which gives the following function of $x$, which has the same sign as the original expression for any fixed $P_1$ and $P_2$:
\begin{equation} \label{xBoundExponential}
    w(x) = ae^{\frac{2(\alpha+\beta)(P_1+P_2)x}{N_0}} + be^{\frac{2(\alpha+\beta)P_1x}{N_0}} + ce^{\frac{2(\alpha+\beta)P_2x}{N_0}} + d,
\end{equation}
where
\begin{equation*}
\begin{aligned}
    a \triangleq \Bar{a}e^{-\frac{\alpha^2(P_1+P_2)^2}{N_0}}, &\qquad b \triangleq \Bar{b} e^{-\frac{(\alpha P_1-\beta P_2)^2}{N_0}}, \\
    c \triangleq\Bar{c}e^{-\frac{(\beta P_1-\alpha P_2)^2}{N_0}}, &\qquad d \triangleq \Bar{d}e^{-\frac{\beta^2(P_1+P_2)^2}{N_0}},
\end{aligned}
\end{equation*}
and 
\begin{equation} \label{barCoeffs}
\begin{aligned}
    \Bar{a} \triangleq p_1p_{11|1} - p_0p_{11|0}, &\qquad \Bar{b} \triangleq p_1p_{10|1} - p_0p_{10|0}, \\
    \Bar{c} \triangleq p_1p_{01|1} - p_0p_{01|0}, &\qquad \Bar{d} \triangleq p_1p_{00|1} - p_0p_{00|0}.
\end{aligned}
\end{equation}
Using~\eqref{xBoundExponential}, we can characterize the decision regions as $\D_0 = \{x\in\mathbb{R} \;|\; w(x) \leq 0\} =\D_1^c$. Note that by the assumptions on $p_1$, $\epsilon_1$ and $\epsilon_2$, we have that $\Bar{d} < 0$, which implies that $w(x)$ is negative as $x\rightarrow -\infty$ for any $P_1$ and $P_2$ (hence the MAP rule detects a 0). Thus, we can completely determine these regions by the boundary points between $\D_0$ and $\D_1$. These boundary points are the same as where $w(x)$ crosses from negative to positive, or vice-versa. Thus, it is relevant to analyze the set $\X = \{x\in\mathbb{R} \;|\; w(x) = 0\}$. Note that applying the results of \cite[Corollary 3.2]{expPolyZeros}, we know that the size of this set is restricted to $|\X| \in \{0,1,2,3\}$. Unfortunately, there is not a general way to solve for the values $x\in\X$ analytically. However, the problem can be split into three cases which can be analyzed without knowing these values explicitly. The decision regions can be expressed as unions of intervals using these boundary points. For example, if $\X = \{x\}$, then $\D_0 = (-\infty, x]$, whereas if $\X = \{x_1,x_2,x_3\}$, with $x_1<x_2<x_3$ then $\D_0 = (-\infty, x_1]\cup[x_2,x_3]$.
\subsection{Case~I: $0 \leq p_1 \leq \frac{\epsilon_1\epsilon_2}{1-\epsilon_1-\epsilon_2+2\epsilon_1\epsilon_2}$}
\begin{proposition} \label{case1nosol}
    In Case~I, there are no real solutions to the equation $w(x)=0$, where $w(x)$ is given in~\eqref{xBoundExponential}. 
\end{proposition} 
\begin{IEEEproof}
    Each of the following inequalities hold due to the condition of Case~I, and $p_1 \leq 0.5$, $0 < \epsilon_1 \leq \epsilon_2 < 0.5$:
\begin{align*}
    \Bar{a} &= p_1p_{11|1} - p_0p_{11|0}\\
    &= p_1(1-\epsilon_1)(1-\epsilon_2) - (1-p_1)\epsilon_1\epsilon_2 \\
    &= (1-\epsilon_1-\epsilon_2+2\epsilon_1\epsilon_2)\bigg(p_1 - \frac{\epsilon_1\epsilon_2}{1-\epsilon_1-\epsilon_2+2\epsilon_1\epsilon_2}\bigg)\\
    &\leq 0 \implies a \leq 0, \\[0.5em] 
    \Bar{b} &= p_1p_{10|1} - p_0p_{10|0}\\
    &= p_1(1-\epsilon_1)\epsilon_2 - (1-p_1)\epsilon_1(1-\epsilon_2) \\
    &= (\epsilon_1+\epsilon_2-2\epsilon_1\epsilon_2)\bigg(p_1 - \frac{\epsilon_1 - \epsilon_1\epsilon_2}{\epsilon_1+\epsilon_2-2\epsilon_1\epsilon_2}\bigg) < 0 \\
    \implies b & < 0,\\[0.5em]
    \Bar{c} &= p_1p_{01|1} - p_0p_{01|0}\\
    &= p_1\epsilon_1(1-\epsilon_2) - (1-p_1)(1-\epsilon_1)\epsilon_2 \\
    &= (\epsilon_1+\epsilon_2-2\epsilon_1\epsilon_2)\bigg(p_1 - \frac{\epsilon_2-\epsilon_1\epsilon_2}{\epsilon_1+\epsilon_2-2\epsilon_1\epsilon_2}\bigg) < 0 \\
    \implies c & < 0, \\[0.5em]
    \Bar{d} &= p_1p_{00|1} - p_0p_{00|0}\\
    &= p_1\epsilon_1\epsilon_2 - (1-p_1)(1-\epsilon_1)(1-\epsilon_2) \\
    &= (1-\epsilon_1-\epsilon_2+2\epsilon_1\epsilon_2)\bigg(p_1 - \frac{1-\epsilon_1-\epsilon_2+\epsilon_1\epsilon_2}{1-\epsilon_1-\epsilon_2+2\epsilon_1\epsilon_2}\bigg)\\
    &< 0 \implies d < 0.
\end{align*}

Thus $\; w(x) < 0 \ \forall x \in \mathbb{R}$ and $w(x)$ has no real roots.
\end{IEEEproof}  
Since there is no decision boundary, no matter what the received signal is, the optimal detection will always be $\hat{x} = 0$. Hence, the error probability is only dependent on the source probability $p_1$. In this case, the senors are not able to send any useful data, so they should not send anything at all. We conclude that the optimal power allocation and corresponding error performance are expressed as follows:
\begin{align*}
    P_1^{* \text{Case~I}} = 0, \quad
    P_2^{* \text{Case~I}} = 0, \quad
    P_e^{* \text{Case~I}} = p_1.
\end{align*}

\subsection{Case~II: $\frac{\epsilon_1\epsilon_2}{1-\epsilon_1-\epsilon_2+2\epsilon_1\epsilon_2} < p_1 \leq \frac{\epsilon_1-\epsilon_1\epsilon_2}{\epsilon_1+\epsilon_2-2\epsilon_1\epsilon_2}$}
Using a similar approach as in the proof of Proposition~\ref{case1nosol}, we can show the following properties about the coefficients of \eqref{xBoundExponential} in this case:
\begin{equation*}
    a > 0, \quad b \leq 0, \quad c\leq0, \quad d < 0.
\end{equation*}
\begin{proposition}
    In Case~II, there is exactly one real root to $w(x)$ in \eqref{xBoundExponential} for any $P_1, P_2 > 0$. Further, this root is also a decision boundary between $\D_0$ and $\D_1$. 
\end{proposition}
\begin{IEEEproof}
Let $P_1,P_2 > 0$. First we show that there exists at least one real solution to $w(x) = 0$. We use the fact that $w(x)$ is continuous and that its asymptotic behaviours are:
\begin{equation*}
    \lim_{x \to -\infty} w(x) = d, \quad \lim_{x \to \infty}w(x) = \infty.
\end{equation*}
Since $d<0$ and $w(x)$ is continuous, it must have at least one root. Next we show that $w(x)$ can have at most one root by showing that once it becomes non-negative, the derivative is always strictly positive, so it can never have another zero. Assume $w(x) \geq 0$, we then have:
\begin{align*}
    \frac{dw}{dx} &= \frac{2(\alpha+\beta)}{N_0}\bigg((P_1+P_2)ae^{\frac{2(\alpha+\beta)(P_1+P_2)x}{N_0}}+\\
    &\qquad\qquad\qquad P_1be^{\frac{2(\alpha+\beta)P_1x}{N_0}}+P_2ce^{\frac{2(\alpha+\beta)P_2x}{N_0}}\bigg)\\
    &= \frac{2(\alpha+\beta)}{N_0}\bigg((P_1+P_2)w(x)-P_2be^{\frac{2(\alpha+\beta)P_1x}{N_0}}\\
    &\qquad\qquad\qquad -P_1ce^{\frac{2(\alpha+\beta)P_2x}{N_0}}-(P_1+P_2)d\bigg) 
    > 0.
\end{align*}
This shows that $w(x)$ has exactly one real root, and this root is a boundary point between $\D_0$ and $\D_1$ as desired.
\end{IEEEproof}
Now we can express the error probability at any $P_1,P_2 > 0$ using the following expression:
\begin{equation} \label{Error1Solution}
    P_e(P_1, P_2) = \hspace{-15pt}\sum_{(l,m)\in\{0,1\}^2}\hspace{-15pt}(p_1p_{lm|1} - p_0p_{lm|0})Q\bigg(\frac{a_{lm}-x}{\sigma}\bigg) + p_0p_{lm|0},
\end{equation}
where $Q$ is the Gaussian tail distribution function, defined as
\begin{equation*}
    Q(x) = \frac{1}{\sqrt{2\pi}}\int_x^\infty e^{-\frac{u^2}{2}}du,
\end{equation*}
$a_{lm}$ are the constellation points and $x$ is the root of \eqref{xBoundExponential} corresponding to $P_1$ and $P_2$.

\begin{proposition} \label{upperBoundPe}
    If $P_1,P_2 > 0$ and $x$ is the corresponding unique decision boundary, then for any $\hat{x} \in \mathbb{R}$, the following is an upper bound on the error probability 
    \begin{equation} \label{upperBoundPeEqn}
     P_{e, \hat{x}}^{\text{\tiny UB}}(P_1, P_2) \hspace{-2pt}\triangleq \hspace{-18pt}\sum_{(l,m)\in\{0,1\}^2}\hspace{-16pt}(p_1p_{lm|1} - p_0p_{lm|0})Q\bigg(\hspace{-2pt}\frac{a_{lm}-\Hat{x}}{\sigma}\hspace{-2pt}\bigg) + p_0p_{lm|0}.
    \end{equation}
\end{proposition}
\begin{IEEEproof}
    This expression corresponds to the error probability associated to using a decision boundary $\hat{x}$. Since the true decision boundary, $x$, from the MAP detection rule is optimal, this must be an upper bound.
\end{IEEEproof}

Now we define the following functions of $P_1$ and $P_2$, where $\Bar{a}$, $\Bar{b}$ and $\Bar{c}$ are as defined in \eqref{barCoeffs}.
    \begin{align} \label{KeqCase2}
    K(P_1) &\triangleq \frac{N_0}{2(\alpha+\beta)P_1}\ln\frac{\Bar{a}}{-\Bar{c}} - \frac{\alpha-\beta}{2}P_1,\\ \label{LeqCase2}
    L(P_2) &\triangleq \frac{N_0}{2(\alpha+\beta)P_2}\ln\frac{\Bar{a}}{-\Bar{b}} - \frac{\alpha-\beta}{2}P_2.
    \end{align}
    Note that these are well defined if $\Bar{c} \neq 0$ and $\Bar{b} \neq 0$, respectively.
    \begin{proposition} \label{xInequalitiesCase2}
        For any $P_1,P_2 > 0$ the following two statements are true if $\Bar{b} \neq 0$ and $\Bar{c} \neq 0$, respectively:
        \begin{align} \label{LIneqCase2}
            x \lesseqgtr \alpha P_1 - L(P_2) &\implies ae^{\frac{2(\alpha + \beta)P_2x}{N_0}} + b \lesseqgtr 0, \\ \label{KIneqCase2}
            x \lesseqgtr \alpha P_2 - K(P_1) &\implies ae^{\frac{2(\alpha + \beta)P_1x}{N_0}} + c \lesseqgtr 0,
        \end{align}
        where the symbol $\lesseqgtr$ means that the statements hold for any of the relations $<$, $>$ or $=$, consistently in the each line.
    \end{proposition}
    \begin{IEEEproof}
        See Appendix~\ref{ProofxInequalitiesCase2}.
    \end{IEEEproof}
    \begin{proposition} \label{xBoundCase2}
        In Case~II, for $P_1,P_2>0$, if $x$ is the corresponding root of \eqref{xBoundExponential}, $\Bar{b} \neq 0$ and $\Bar{c} \neq 0$, then the following two inequalities hold:
        \begin{align} \label{LBoundCase2}
            x &> \alpha P_1 - L(P_2), \\ \label{KBoundCase2}
            x &> \alpha P_2 - K(P_1).
        \end{align}
    \end{proposition}
    \begin{IEEEproof}
        Using Proposition~\ref{xInequalitiesCase2}, we have
        \begin{align*}
            x \leq \alpha P_1 - L(P_2) \implies \hspace{-1pt}ae^{\frac{2(\alpha + \beta)P_2x}{N_0}} + b \leq 0 \implies w(x) < 0,\\
            x \leq \alpha P_2 - K(P_1)\hspace{-3pt}\implies \hspace{-1pt}ae^{\frac{2(\alpha + \beta)P_1x}{N_0}} + c \leq 0 \implies w(x) < 0.
        \end{align*}
        Hence, $x$ could not be a zero of \eqref{xBoundExponential} in either of these cases, showing the desired result.
    \end{IEEEproof}

\begin{proposition} \label{PropCase2P1XInfBound}
    In Case~II, let $\Bar{P}_1$, $P_1'$, $\Bar{P}_2$ and $P_2'$ be arbitrary real numbers such that $0<\Bar{P}_1<P_1'$, $0<\Bar{P}_2<P_2'$. Then the root of \eqref{xBoundExponential}, $x$,  satisfies the following two inequalities as a function of $P_2$ and $P_1$, respectively:
    \begin{align} \label{Case2P1XInfBound}
        \inf_{P_1 \in [\Bar{P}_1,P_1']} x - \alpha P_1 + L(P_2) &> 0, \quad\text{for } \Bar{b} \neq 0, P_2>0,\\
        \label{Case2P2XInfBound}
        \inf_{P_2 \in [\Bar{P}_2,P_2']} x - \alpha P_2 + K(P_1) &> 0, \quad\text{for } \Bar{c} \neq 0, P_1>0.
    \end{align}
\end{proposition}
\begin{IEEEproof}
    See Appendix~\ref{P1PrimeIneqProof}.
\end{IEEEproof}

\begin{theorem} \label{Case2DecreaseP1P2}
    In Case~II, $P_e(P_1,P_2)$ is decreasing in $P_1$ and $P_2$ for $P_1,P_2 > 0$.
\end{theorem}

\begin{IEEEproof}
    We will show the following two statements:
    \begin{enumerate}
        \item If $0<P_1<P_1'$, $0<P_2$, then $P_e(P_1, P_2) > P_e(P_1', P_2)$.
        \item If $0<P_1$, $0<P_2<P_2'$, then $P_e(P_1, P_2) > P_e(P_1, P_2')$.
    \end{enumerate}
    
    To show Statement 1, fix $0<P_1<P_1'$, $0<P_2$. Let $x$ and $x'$ be the roots of \eqref{xBoundExponential} corresponding to the pairs ($P_1,P_2)$ and $(P_1',P_2)$, respectively. We define the following sequence $\{P_{1,i}\}_{i=0}^\infty$ recursively.
    \begin{equation*}
        P_{1,0} = P_1, \qquad\qquad P_{1,i+1} = \begin{cases}
            P_1',&\Bar{b} = 0\\
            \frac{1}{\alpha}\big(x_i+L(P_2)\big),&\Bar{b} \neq 0
        \end{cases}
    \end{equation*}
    where $x_i$ is the root to \eqref{xBoundExponential} corresponding to $P_{1,i}$, $\Bar{b}$ is from \eqref{barCoeffs} and $L$ is as defined in \eqref{LeqCase2}. Note that if $\Bar{b} \neq 0$ and $P_{1,i} < P_1'$, applying \eqref{Case2P1XInfBound} gives 
    \begin{align*}
        P_{1,i+1} - P_{1,i} &= \frac{1}{\alpha}\big(x_i+L(P_2)\big) - P_{1,i} \\
        &\geq \frac{1}{\alpha}\inf_{\Bar{P}_1 \in [P_1,P_1']} \Bar{x} - \alpha \Bar{P}_1 + L(P_2) \stackrel{\eqref{Case2P1XInfBound}}{>} 0,
    \end{align*}
    where $\Bar{x}$ denotes the root of \eqref{xBoundExponential} for $(\Bar{P}_1,P_2)$. This shows that the sequence $\{P_{1,i}\}_{i=0}^\infty$ increases by at least this constant if $P_{1,i} < P_1'$. Therefore there exists $i'$ large enough such that $P_{1,i'} \geq P_1'$. Hence, it is sufficient to show that for all $i$ 
    \begin{equation*}
        P_e(P_{1,i}, P_2) - P_e(P_{1,i+1}, P_2) > 0.
    \end{equation*}
    Using the upper bound in Proposition~\ref{upperBoundPe}, it is sufficient to show 
    \begin{align*}
        P_e(P_{1,i}, P_2) - P_{e,x_i}^{\text{\tiny UB}}(P_{1,i+1}, P_2) &> 0 \\
        \iff P_{e,x_i}^{\text{\tiny UB}}(P_{1,i}, P_2) - P_{e,x_i}^{\text{\tiny UB}}(P_{1,i+1}, P_2) &> 0,
    \end{align*}
    since $P_e(P_{1,i}, P_2) = P_{e,x_i}^{\text{\tiny UB}}(P_{1,i}, P_2)$ by its definition. 
    
    It is now sufficient to show $P_{e,x_i}^{\text{\tiny UB}}$ is decreasing in $P_1$ over $(P_{1,i}, P_{1,i+1})$. The derivative of this expression is:
    \begin{align} \label{PUBderivativeP1}
    \nonumber
        \frac{dP_{e,x_i}^{\text{\tiny UB}}}{dP_1} &=
        \frac{-1}{\sigma\sqrt{2\pi}}e^{\frac{-x_i^2}{N_0}}\Bigg(\alpha\bigg(ae^{\frac{2\alpha(P_1+P_2)x_i}{N_0}} + be^{\frac{2(\alpha P_1 -\beta P_2)x_i}{N_0}}\bigg) \\
        &\qquad-\beta\bigg(ce^{\frac{2(-\beta P_1 + \alpha P_2)x_i}{N_0}} + de^{\frac{-2\beta(P_1+P_2)x_i}{N_0}}\bigg)\Bigg).
    \end{align}
    If $\Bar{b} = 0$, this derivative is negative for any $P_1$ (since this is equivalent to $b=0$). If $\Bar{b} \neq 0$, then we apply \eqref{LIneqCase2} from Proposition~\ref{xInequalitiesCase2} and conclude that
    \begin{align*}
        &P_1 < P_{1,i+1} = \frac{1}{\alpha}\big(x_i+L(P_2)\big) \\
        &\iff  x_i > \alpha P_1 - L(P_2) \\
        &\implies ae^{\frac{2(\alpha + \beta)P_2x_i}{N_0}} + b > 0 \\
        &\implies ae^{\frac{2\alpha(P_1+P_2)x_i}{N_0}} + be^{\frac{2(\alpha P_1 -\beta P_2)x_i}{N_0}} > 0 \\
        &\implies \frac{dP_{e,x_i}^{\text{\tiny UB}}}{dP_1} < 0.
    \end{align*}
    The proof of Statement 2 is omitted as it follows the exact same steps as above, replacing the roles of $L(P_2)$ with $K(P_1)$, $\Bar{b}$ with $\Bar{c}$, and applying \eqref{KIneqCase2} and \eqref{Case2P2XInfBound} instead of \eqref{LIneqCase2} and \eqref{Case2P1XInfBound}.    
\end{IEEEproof}

Thus the optimal power allocation and corresponding error performance can be expressed as follows, where $p_{lm|i}$ are as given in \eqref{conditionalProbs}, $x^*$ is the root to \eqref{xBoundExponential} and $a_{lm}^*$ are the constellation points corresponding to $P_1^*$ and $P_2^*$.
\begin{align} \label{PeCase2}
    \nonumber
    &P_1^{* \text{Case~II}} = \sqrt{P_1^{\text{max}}}, \quad
    P_2^{* \text{Case~II}} = \sqrt{P_2^{\text{max}}}, \\
    P_e^{* \text{Case~II}}\hspace{-2pt} &= \hspace{-17pt}\sum_{(l,m)\in\{0,1\}^2}\hspace{-13pt}(p_1p_{lm|1} - p_0p_{lm|0})Q\bigg(\frac{a_{lm}^*-x^*}{\sigma}\bigg) + p_0p_{lm|0}.
\end{align}

\subsection{Case~III: $\frac{\epsilon_1-\epsilon_1\epsilon_2}{\epsilon_1+\epsilon_2-2\epsilon_1\epsilon_2} < p_1 \leq 0.5$}
First note that the condition of this case implies $\epsilon_1 \neq \epsilon_2$. Also using the same reasoning as in the proof of Proposition~\ref{case1nosol}, we make the following observations about the coefficients of $w(x)$ in~\eqref{xBoundExponential}:
\begin{equation*}
    a > 0, \quad b > 0, \quad c<0, \quad d < 0.
\end{equation*}
We define the following functions of $P_1$, where $\Bar{a}$, $\Bar{b}$, $\Bar{c}$ and $\Bar{d}$ are as defined in \eqref{barCoeffs}:
\begin{align} \label{nonUniformP2Tilde}
    \Tilde{P}_2(P_1) &\triangleq  \frac{N_0}{2(\alpha+\beta)^2P_1}\ln\frac{\Bar{a}\Bar{d}}{\Bar{b}\Bar{c}},
    \\ \label{KalphaDef}
    K_\alpha (P_1) &\triangleq \frac{N_0}{2(\alpha+\beta)P_1}\ln\frac{\Bar{a}}{-\Bar{c}} - \frac{\alpha-\beta}{2}P_1, \\ \label{KbetaDef}
    K_\beta (P_1) &\triangleq \frac{N_0}{2(\alpha+\beta)P_1}\ln\frac{-\Bar{d}}{\Bar{b}} + \frac{\alpha-\beta}{2}P_1.
\end{align}
The condition for Case~III combined with the other assumptions on the problem's parameters ensure that these functions are real valued for all $P_1>0$. Also note that expanding \eqref{nonUniformP2Tilde} at $P_1 = \sqrt{P_1^{\text{max}}}$ gives the expression in \eqref{nonUniformP2TildeMax}.

\begin{proposition} \label{xInequalitiesCase3}
    For any $P_1,P_2>0$ the following two statements are true:
    \begin{align} \label{KalphaIneq}
        x \lesseqgtr \alpha P_2 - K_\alpha (P_1) &\implies ae^{\frac{2(\alpha+\beta)P_1x}{N_0}} + c\lesseqgtr 0, \\ \label{KbetaIneq}
        x \lesseqgtr -\beta P_2 + K_\beta (P_1) &\implies be^{\frac{2(\alpha+\beta)P_1x}{N_0}} + d \lesseqgtr 0.
    \end{align}
    \begin{IEEEproof}
        These statements follow directly from rearranging these equations and using the definitions of $K_\alpha$ and $K_\beta$ in \eqref{KalphaDef} and \eqref{KbetaDef}, respectively. The steps are the same as in the proof of Proposition~\ref{xInequalitiesCase2}.
    \end{IEEEproof}
\end{proposition}

\begin{proposition} \label{xBoundsP2Case3}
    In Case~III, for any $P_1,P_2>0$, if $x$ is a corresponding root of \eqref{xBoundExponential}, then it must satisfy
    \begin{equation} \label{xBoundsP2Case3Eqn}
    \begin{cases} 
      x \in \big(\alpha P_2 - K_\alpha (P_1), -\beta P_2 + K_\beta (P_1)\big),& P_2 < \Tilde{P}_2(P_1)\\
      x = \alpha P_2 - K_\alpha (P_1) = -\beta P_2 + K_\beta (P_1), & P_2 = \Tilde{P}_2(P_1)\\
      x \in \big(-\beta P_2 + K_\beta (P_1), \alpha P_2 -K_\alpha (P_1)\big), & P_2 > \Tilde{P}_2(P_1)
    \end{cases}
    \end{equation}
\end{proposition}
\begin{IEEEproof}
    First we note that these intervals are valid and have non-zero length since
    \begin{align*}
         &\quad -\beta P_2 + K_\beta (P_1) - \big(\alpha P_2 - K_\alpha (P_1)\big)\\
         &= K_\alpha(P_1) + K_\beta(P_1) - (\alpha + \beta)P_2\\
         &= (\alpha + \beta)\big(\Tilde{P}_2(P_1) - P_2\big),
    \end{align*}
    so
    \begin{equation*}
         P_2 \lessgtr \Tilde{P}_2(P_1) \implies \alpha P_2 - K_\alpha (P_1) \lessgtr -\beta P_2 + K_\beta (P_1).
    \end{equation*}
    If $x$ is outside these intervals, then we apply \eqref{KalphaIneq} and \eqref{KbetaIneq} from Proposition~\ref{xInequalitiesCase3} to see that  $w(x) \neq 0$, so $x$ could not be a root of \eqref{xBoundExponential}.
\end{IEEEproof}

\begin{figure}[hbtp]
\centerline{\includegraphics[scale=0.45]{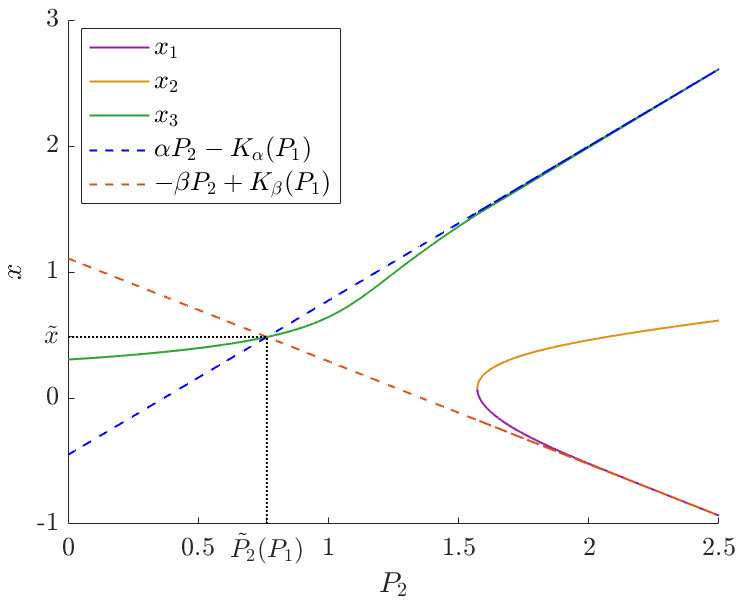}}
\caption{The roots of \eqref{xBoundExponential}, $x_1$, $x_2$ and $x_3$, as a function of $P_2$ in Case~III ($p_1 = 0.4$, $\epsilon_1 = 0.01$, $\epsilon_2 = 0.05$, $N_0 = 1$, $P_1 = 1$).}
\label{case3solfig}
\end{figure}

\begin{proposition} \label{case3OneSolP2}
    In Case~III, for any $P_1,P_2>0$, there will be one or three boundary points between $\D_0$ and $\D_1$. Further, if $P_2 \in [0,\Tilde{P}_2(P_1)]$ there is exactly one boundary point between $\D_0$ and $\D_1$. 
\end{proposition}
\begin{IEEEproof}
    Let $P_1,P_2>0$. We have the following asymptotic behaviours of $w(x)$ in \eqref{xBoundExponential}:
\begin{equation*}
    \lim_{x \to -\infty} w(x) = d, \quad \lim_{x \to \infty}w(x) = \infty.
\end{equation*}
Since $d<0$, and $w$ is continuous, we conclude that there must be an odd number of crossing points. Combing this fact with~\cite[Corollary 3.2]{expPolyZeros} yields that there are one or three crossing points. Now let $P_2 \in [0,\Tilde{P}_2(P_1)]$. We assume $w(x) = 0$, and perform the following derivative analysis:
\begin{align*}
    \frac{dw}{dx} &= \frac{2(\alpha+\beta)}{N_0}\bigg((P_1+P_2)ae^{\frac{2(\alpha+\beta)(P_1+P_2)x}{N_0}}\\
    &\qquad\qquad\qquad+P_1be^{\frac{2(\alpha+\beta)P_1x}{N_0}}+P_2ce^{\frac{2(\alpha+\beta)P_2x}{N_0}}\bigg)\\
    &= \frac{2(\alpha+\beta)}{N_0}\bigg((P_1+P_2)w(x)-P_2be^{\frac{2(\alpha+\beta)P_1x}{N_0}}\\
    &\qquad\qquad\qquad-P_1ce^{\frac{2(\alpha+\beta)P_2x}{N_0}}-(P_1+P_2)d\bigg) \\ 
    &= \frac{-2(\alpha+\beta)}{N_0}\bigg(P_2(be^{\frac{2(\alpha+\beta)P_1x}{N_0}}+d)\\
    &\qquad\qquad\qquad+P_1(ce^{\frac{2(\alpha+\beta)P_2x}{N_0}}+d)\bigg) \\
    &> 0.
\end{align*}
This last inequality is true since Propositions~\ref{xInequalitiesCase3} and \ref{xBoundsP2Case3} imply 
\begin{equation*}
    be^{\frac{2(\alpha+\beta)P_1x}{N_0}}+d \leq 0.
\end{equation*}
Therefore $w(x)$ is strictly increasing at any zero, and hence must only have one root, and this root must be a boundary point between $\D_0$ and $\D_1$ as desired.
\end{IEEEproof}
\begin{proposition} \label{Case3P2XInfBoundProp}
    In Case~III, for any $P_1>0$, let $\Bar{P}_2$ and $P_2'$ be arbitrary real numbers such that $0<\Bar{P}_2<P_2'<\Tilde{P}_2(P_1)$, where $\Tilde{P}_2(P_1)$ is as defined in \eqref{nonUniformP2Tilde}. Then the root of \eqref{xBoundExponential}, $x$,  satisfies the following two inequalities:
    \begin{align} \label{Case3KaXInfBound}
        \inf_{P_2 \in [\Bar{P}_2,P_2']} x - \alpha P_2 + K_\alpha(P_1) &> 0,\\
        \label{Case3KbXInfBound}
        \inf_{P_2 \in [\Bar{P}_2,P_2']} K_\beta(P_1)-\beta P_2 - x &> 0.
    \end{align}
\end{proposition}
\begin{IEEEproof}
The details are omitted as it follows the same steps as the proof of Proposition~\ref{PropCase2P1XInfBound}.
\end{IEEEproof}
The results about the roots of \eqref{xBoundExponential} in Case~III are illustrated in Fig.~\ref{case3solfig}.

Now we define the following two functions of $x$, for any fixed $P_1$ and $P_2$:
\begin{align*}
    g(x) &\triangleq (p_1p_{11|1} - p_0p_{11|0})Q\bigg(\frac{\alpha P_1 + \alpha P_2 -x}{\sigma}\bigg) \nonumber \\
    &\quad+ (p_1p_{01|1} - p_0p_{01|0})Q\bigg(\frac{-\beta P_1 + \alpha P_2 -x}{\sigma}\bigg),
\end{align*}
\vspace{-2.5pt}
\begin{align*}
    h(x) &\triangleq (p_1p_{10|1} - p_0p_{10|0})Q\bigg(\frac{\alpha P_1 - \beta P_2 -x}{\sigma}\bigg) \nonumber \\
    &\quad+ (p_1p_{00|1} - p_0p_{00|0})Q\bigg(\frac{-\beta P_1 - \beta P_2 -x}{\sigma}\bigg).
\end{align*}
\begin{proposition} \label{FGminimization}
    In Case~III, for any $P_1,P_2>0$, $g(x)$ is minimized at $x=\alpha P_2 - K_\alpha(P_1)$ and $h(x)$ is minimized at $x = -\beta P_2 + K_\beta(P_1)$.
\end{proposition}
\begin{IEEEproof}
    See Appendix~\ref{ProofFGMinimization}.
\end{IEEEproof}
\begin{theorem} \label{Case3P2ErrorMinimization}
    In Case~III, $P_e(P_1,\Tilde{P}_2(P_1)) < P_e(P_1, P_2)$, $ \forall P_1>0, P_2 \neq \Tilde{P}_2(P_1)$.
\end{theorem}

\begin{IEEEproof}
    Fix $P_1>0$. Let $\Tilde{a}_{lm}$ be the constellation points and $\Tilde{x}$ be the root of \eqref{xBoundExponential} corresponding to $P_1$ and $\Tilde{P}_2(P_1)$. Let $P_2 \neq \Tilde{P_2}(P_1)$. Let $a_{lm}$ be the constellation points and $\X$ denote the set of roots of \eqref{xBoundExponential} corresponding to $P_1$ and $P_2$. First we analyze the case where $|\X| = 1$. Let $\X = \{x\}$. In this case, the error expression, $P_e(P_1, P_2)$, will have the same form as given in \eqref{Error1Solution}. Also, since there is a unique root, $\Tilde{x}$, at $\Tilde{P_2}(P_1)$, $P_e(P_1,\Tilde{P}_2(P_1))$ takes the form of \eqref{Error1Solution} as well. Therefore, we must show
\begin{align*}
    &\sum_{(l,m)\in\{0,1\}^2}\hspace{-10pt}(p_1p_{lm|1} - p_0p_{lm|0})Q\bigg(\frac{\Tilde{a}_{lm}-\Tilde{x}}{\sigma}\bigg) <\\
    &\hspace{65pt}\sum_{(l,m)\in\{0,1\}^2}\hspace{-10pt}(p_1p_{lm|1} - p_0p_{lm|0})Q\bigg(\frac{a_{lm}-x}{\sigma}\bigg) \\
    &\iff g\big(\alpha P_2 - K_\alpha(P_1)\big) + h\big(-\beta P_2 + K_\beta(P_1)\big) <\\
    &\hspace{190pt}g(x) + h(x).
\end{align*}

This result follows immediately by applying Proposition~\ref{FGminimization}, since $x \neq \alpha P_2 - K_\alpha(P_1)$ and $x \neq -\beta P_2 + K_\beta(P_1)$ for $P_2 \neq \Tilde{P_2}(P_1)$ from Proposition~\ref{xBoundsP2Case3}.

For the case that $|\X| = 3$, we represent this as $\X = \{x_1, x_2, x_3\}$ such that $x_1 < x_2 < x_3$. Further we note that the decision regions are represented in terms of these boundaries as $\D_1 = (x_1,x_2)\cup(x_3,\infty)$. This gives rise to the following expression for the error probability, also noting that it can be expressed in terms of $g$ and $h$:
\begin{align} \nonumber \label{error3Solution}
    P_e(P_1,P_2) &= \hspace{-10pt}\sum_{(l,m)\in\{0,1\}^2}\hspace{-10pt}(p_1p_{lm|1} - p_0p_{lm|0})\Bigg(Q\bigg(\frac{a_{lm}-x_1}{\sigma}\bigg)\\
    &- Q\bigg(\frac{a_{lm}-x_2}{\sigma}\bigg) + Q\bigg(\frac{a_{lm}-x_3}{\sigma}\bigg)\Bigg) + p_0p_{lm|0}, \\ \nonumber
    &= g(x_1) - g(x_2) + g(x_3) \\ \nonumber
    &\quad+ h(x_1) - h(x_2) + h(x_3)+ \hspace{-10pt}\sum_{(l,m)\in\{0,1\}^2}\hspace{-10pt}p_0p_{lm|0}.
\end{align}
Hence we must show
\begin{align} \label{3solIneqality}
     &g\big(\alpha P_2 - K_\alpha(P_1)\big) + h\big(-\beta P_2 + K_\beta(P_1)\big) < \nonumber \\
     &\quad g(x_1) - g(x_2) + g(x_3) + h(x_1) - h(x_2) + h(x_3).
\end{align}
First note that by applying Proposition~\ref{FGminimization}, we have
\begin{equation} \label{case3P2PartialIneq}
    g\big(\alpha P_2 - K_\alpha(P_1)\big) + h\big(-\beta P_2 + K_\beta(P_1)\big) < g(x_3) + h(x_1).
\end{equation}
Now we will show 
\begin{equation} \label{case3P2PositivePart}
g(x_1) - g(x_2) + h(x_3) - h(x_2) \geq 0.
\end{equation}
Since $|\X| = 3$ we can infer that $P_2 > \Tilde{P}_2(P_1)$ by taking the contra-positive of Proposition~\ref{case3OneSolP2}. Then we can apply Proposition~\ref{xBoundsP2Case3} which implies $x_2 \in (-\beta P_2 + K_\beta (P_1), \alpha P_2 -K_\alpha (P_1))$. Finally, applying the same derivative analysis as in Proposition~\ref{FGminimization} shows that $g(x)$ is decreasing on $[x_1,x_2]$ and $h(x)$ is increasing on $[x_2,x_3]$. This implies \eqref{case3P2PositivePart}.
Combining \eqref{case3P2PartialIneq} with \eqref{case3P2PositivePart} implies \eqref{3solIneqality}.
\end{IEEEproof}
\begin{theorem} \label{Case3P2decreasingError}
    In Case~III, if $0<P_1$, $0<P_2<P_2'<\Tilde{P}_2(P_1)$, then $P_e(P_1, P_2) > P_e(P_1, P_2')$.
\end{theorem}
\begin{IEEEproof}
    Fix $0<P_1$, $0<P_2<P_2'<\Tilde{P}_2$. Let $x$ and $x'$ be the roots of \eqref{xBoundExponential} corresponding to $P_2$ and $P_2'$, respectively. We define the following sequence $\{P_{2,i}\}_{i=0}^\infty$ recursively.
    \begin{align*}
        P_{2,0} &= P_2,\\ 
        P_{2,i+1} &= \min\bigg(\frac{1}{\alpha}\big(x_i+K_\alpha(P_1)\big), \frac{1}{\beta}\big(K_\beta(P_1)-x_i\big)\bigg),
    \end{align*}
    where $x_i$ is the root of \eqref{xBoundExponential} corresponding to $P_{2,i}$. Note that \eqref{xBoundsP2Case3Eqn} implies $P_{2,i} \leq \Tilde{P}_2$ for any $i\geq 0$, so there is always a unique $x_i$. If $P_{2,i} < P_2'$, applying \eqref{Case3KaXInfBound} and \eqref{Case3KbXInfBound} means one of the following two statements must be true:
    \begin{align*}
        P_{2,i+1}-P_{2,i} &= \frac{1}{\alpha}\big(x_i+K_\alpha(P_1)\big) - P_{2,i}\\
        &\geq\underbrace{\frac{1}{\alpha}\inf_{\Bar{P}_2 \in [P_2,P_2']} \Bar{x} - \alpha \Bar{P}_2 + K_\alpha(P_1)}_{\triangleq K_\alpha'} \stackrel{\eqref{Case3KaXInfBound}}{>} 0,
    \end{align*}
    or
    \begin{align*}
        P_{2,i+1}-P_{2,i} &= \frac{1}{\beta}\big(K_\beta(P_1)-x_1\big) - P_{2,i}\\
        &\geq\underbrace{\frac{1}{\beta}\inf_{\Bar{P}_2 \in [P_2,P_2']} K_\beta(P_1)-\beta \Bar{P}_2 - \Bar{x}}_{\triangleq K_\beta'} \stackrel{\eqref{Case3KbXInfBound}}{>} 0,
    \end{align*}
    so
    \begin{align*}
        P_{2,i+1}-P_{2,i} \geq \min(K_\alpha', K_\beta') 
        > 0.
    \end{align*}
    where $\Bar{x}$ denotes the root of \eqref{xBoundExponential} for $(P_1,\Bar{P}_2)$. This means that the sequence $\{P_{2,i}\}_{i=0}^\infty$ increases by at least this fixed constant if $P_{2,i} < P_2'$. Therefore, there exists $i'$ large enough such that $P_{2,i'} \geq P_2'$. Hence, it is sufficient to show that for all $i$ 
    \begin{equation*}
        P_e(P_1, P_{2,i}) - P_e(P_1, P_{2,i+1}) > 0.
    \end{equation*}
    Using the upper bound in Proposition~\ref{upperBoundPe} (which is valid for exactly the same reasons as before), it is sufficient to show 
    \begin{align*}
        P_e(P_1, P_{2,i}) - P_{e,x_i}^{\text{\tiny UB}}(P_1, P_{2,i+1}) &> 0 \\
        \iff P_{e,x_i}^{\text{\tiny UB}}(P_1, P_{2,i}) - P_{e,x_i}^{\text{\tiny UB}}(P_1, P_{2,i+1}) &> 0,
    \end{align*}
    since $P_e(P_1, P_{2,i}) = P_{e,x_i}^{\text{\tiny UB}}(P_1, P_{2,i})$ by definition in \eqref{upperBoundPeEqn}. It is now sufficient to show $P_{e,x_i}^{\text{\tiny UB}}$ is decreasing in $P_2$ over $(P_{2,i}, P_{2,i+1})$. We have
    \begin{align*} 
        \frac{dP_{e,x_i}^{\text{\tiny UB}}}{dP_2} &= \frac{-1}{\sigma\sqrt{2\pi}}e^{\frac{-x_i^2}{N_0}}\Bigg(\alpha\bigg(ae^{\frac{2\alpha(P_1+P_2)x_i}{N_0}} + ce^{\frac{2(-\beta P_1 + \alpha P_2)x_i}{N_0}}\bigg) \nonumber\\
        &\qquad-\beta\bigg(be^{\frac{2(\alpha P_1 - \beta P_2)x_i}{N_0}}+ de^{\frac{-2\beta(P_1+P_2)x_i}{N_0}}\bigg)\Bigg),
    \end{align*} 
   where applying Proposition~\ref{xInequalitiesCase3} gives
    \begin{align*}
        P_2 <&\; P_{2,i+1} \leq \frac{1}{\alpha}\big(x_i + K_\alpha(P_1)\big)\\
        &\implies x_i > \alpha P_2 - K_\alpha(P_1) \\
        &\implies\; 0 < ae^{\frac{2(\alpha+\beta)P_1x_i}{N_0}} + c \\
        &\implies\; 0 < ae^{\frac{2\alpha(P_1+P_2)x_i}{N_0}} + ce^{\frac{2(-\beta P_1 + \alpha P_2)x_i}{N_0}},
    \end{align*}
    and
    \begin{align*}
        P_2 <&\; P_{2,i+1} \leq \frac{1}{\beta}\big(K_\beta(P_1)-x_i\big)\\
        &\implies x_i < -\beta P_2 + K_\beta(P_1) \\
        &\implies\; 0 > be^{\frac{2(\alpha+\beta)P_1x_i}{N_0}} + d \\
        &\implies\; 0 > be^{\frac{2(\alpha P_1 - \beta P_2)x_i}{N_0}} + de^{\frac{-2\beta(P_1+P_2)x_i}{N_0}},
    \end{align*}
    so
    \begin{equation*}
        \implies \frac{dP_{e,x_i}^{\text{\tiny UB}}}{dP_2} < 0.
    \end{equation*}
\end{IEEEproof}
Combining Theorems~\ref{Case3P2ErrorMinimization} and \ref{Case3P2decreasingError} we can conclude that in Case~III, the optimal power allocation for $P_2$ given $P_1>0$ is
\begin{equation*}
    P_2^*(P_1) = \min\big(\sqrt{P_2^{\text{max}}}, \Tilde{P}_2(P_1)\big)
\end{equation*}
Now that we have the optimal $P_2$ allocation for a fixed $P_1$ we can analyze the $P_1$ optimization using this result.

\begin{theorem} \label{Case3DecreaseP1}
    In Case~III, $P_e\big(P_1,P_2^*(P_1)\big)$ is decreasing in $P_1$, for all $P_1>0$.
\end{theorem}

\begin{IEEEproof}
    See Appendix~\ref{ProofCase3DecreaseP1}.
\end{IEEEproof}

Hence the optimal power allocation and corresponding error performance can be expressed as follows, where $x^*$ is the root of \eqref{xBoundExponential}, and $a_{lm}^*$ are the corresponding constellation points to $P_1^*$ and $P_2^*$:
\begin{align} \label{PeCase3}
    \nonumber
    P_1^{* \text{Case~III}} &= \sqrt{P_1^{\text{max}}},\\ \nonumber
    P_2^{* \text{Case~III}} &= \min\big(\sqrt{P_2^{\text{max}}}, \Tilde{P}_2(\sqrt{P_1^{\text{max}}})\big),\\
    P_e^{* \text{Case~III}} &= \hspace{-19pt}\sum_{(l,m)\in\{0,1\}^2}\hspace{-16pt}(p_1p_{lm|1} - p_0p_{lm|0})Q\bigg(\frac{a_{lm}^*-x^*}{\sigma}\bigg) + p_0p_{lm|0}.
\end{align}
\subsection{High SNR Behaviour}
For this analysis it is defined that high SNR means $N_0 \rightarrow 0$. This is a reasonable assumption since each sensor's SNR should be growing at similar rates, and one sensor should not have infinitely more power than the other. In Case~I, there is nothing to consider, since the error probability is constant. In Cases II and III, the high SNR behaviour can be analyzed by considering a system that knows which point in the constellation $\C$ was sent. Note this is equivalent to knowing $X_1$ and $X_2$ perfectly, except if two constellation points overlap, i.e., $a_{ij} = a_{i^{'}j^{'}}$ for some $i \neq i^{'}$, $j \neq j^{'}$. Also note that the high SNR behaviour of only sending Sensor $i$ is always $\epsilon_i$, $i \in \{1,2\}$. In the case that both sensor values are used, the following is the MAP detection rule for knowing $X_1=x_1$ and $X_2=x_2$:
\begin{align*} 
    \hat{x}(x_1,x_2) &= \underset{i\in\{0,1\}}{\arg\max}\; \text{Pr}(X = i\;|\;X_1=x_1, X_2=x_2) \\
   & = \underset{i\in\{0,1\}}{\arg\max}\;p_ip_{x_1x_2|i}.
\end{align*}
The decision conditions can be expressed in terms of the constants defined in \eqref{xBoundExponential} as follows:
\begin{align*}
    \Bar{a} > 0 &\iff \Hat{x}
    (1,1) = 1, \qquad\qquad\Bar{b} > 0 \iff \Hat{x}
    (1,0) = 1, \\
    \Bar{c} > 0 &\iff \Hat{x}
    (0,1) = 1, \qquad\qquad\Bar{d} > 0 \iff \Hat{x}
    (0,0) = 1.
\end{align*}
\subsubsection{Case~II}
Based on the values of $\Bar{a},\Bar{b},\Bar{c},\Bar{d}$ in this case, we have the following detection rule:
\begin{align*}
    \Hat{x}
    (1,1) &= 1, \qquad\qquad \Hat{x}
    (1,0) = 0, \\
    \Hat{x}
    (0,1) &= 0, \qquad\qquad \Hat{x}
    (0,0) = 0.
\end{align*}
Since $\Hat{x}(1,0) = \Hat{x}(0,1)$, it does not matter if these two constellation points overlap (and these are the only two constellation points which can possibly overlap). Finally, the high SNR behaviour is calculated to be
\begin{align} \label{case2HighSNR}
    \lim_{N_0 \rightarrow 0}P_e^*(P_1^*,P_2^*) &= p_1(p_{00|1}+p_{01|1}+p_{10|1})+p_0p_{11|0} \nonumber \\
    &= \epsilon_1\epsilon_2 + p_1(\epsilon_1 +  \epsilon_2 - 2\epsilon_1\epsilon_2).
\end{align}
\subsubsection{Case~III}
In this case, there are two interesting cases to consider. First, for the optimal allocation $P_2 = P_2^*$, we have that  $P_2^* \rightarrow 0$ for high SNR; so the error performance in this case approaches the performance of only sending Sensor 1, which is $\epsilon_1$. In the alternative case that both sensors use all their power, the detection rule is as follows:
\begin{align*}
    \Hat{x}
    (1,1) &= 1, \qquad\qquad \Hat{x}
    (1,0) = 1, \\
    \Hat{x}
    (0,1) &= 0, \qquad\qquad \Hat{x}
    (0,0) = 0.
\end{align*}
If no constellation points overlap, $\hat{X} = X_1$, which implies 
\begin{equation*}
    \lim_{N_0 \rightarrow 0}P_e(\sqrt{P_1^{\text{max}}},\sqrt{P_2^{\text{max}}}) = \epsilon_1, \qquad P_1^{\text{max}}\neq P_2^{\text{max}}.
\end{equation*}
However, if $P_1^{\text{max}} = P_2^{\text{max}}=P^{\text{max}}$ for some $P^{\text{max}} > 0$, then $a_{01}=a_{10}$. Let $P_e^i$, $i \in \{0,1\}$ be the error probability if we decide to detect $i$ for 10/01. Since $p_1 \leq p_0$ and
\begin{align*}
    P_e^0 &= \epsilon_1\epsilon_2 + p_1(\epsilon_1 +  \epsilon_2 - 2\epsilon_1\epsilon_2)\\
    P_e^1 &=
    \epsilon_1\epsilon_2 + p_0(\epsilon_1 +  \epsilon_2 - 2\epsilon_1\epsilon_2)\\
    \implies P_e^0 &\leq P_e^1,
\end{align*}
we conclude to decide 0, and the final expression for the high SNR behaviour is
\begin{equation*}
    \lim_{N_0 \rightarrow 0}P_e\big(\sqrt{P^{\text{max}}},\sqrt{P^{\text{max}}}\big) = \epsilon_1\epsilon_2 + p_1(\epsilon_1 +  \epsilon_2 - 2\epsilon_1\epsilon_2).
\end{equation*}
These results are demonstrated in Fig. \ref{case3errorVarySNR}, where the curve for using both sensors at their max power has a larger end behaviour than the derived optimal constellation design. Note that in Case~II, $P_e^0 < \epsilon_1$, but in Case~III, $\epsilon_1 < P_e^0 < \epsilon_2$.
\section{Numerical and Simulation Results} \label{numericalSect}
In this section, we illustrate the results of this paper numerically for specific parameter sets of the problem setup. We show that the theoretical results proven in the previous section are also supported by simulated experiments. In what follows, the SNR is defined as the geometric average of the available power allocations, reported in dB (i.e., $\text{SNR (dB)} = 10\log_{10}(\text{SNR})$): 
\begin{equation} \label{snrDef}
    \text{SNR}^{\text{max}} \triangleq \frac{\sqrt{P_1^{\text{max}}P_2^{\text{max}}}}{N_0}.
\end{equation}
Even though Sensor 2 does not necessarily use all of its allocated power, defining the SNR this way is sensible because the sensors have independent power constraints. If the sensors had a joint power constraint, it would be more appropriate to use the true SNR.
\subsection{Simulated Validation of Main Results}
The experimental data is produced by sending 500,000 independent source bits via two simulated sensors and MAC, then using the MAP detection rule given in \eqref{MAPDecodeExpression}, the error probability is calculated. We will show in two ways that the simulations overlap with the theoretical results. First we show that the minimization problem is solved at the correct value of $P_2$ in Case~III. Then we show that the error probability when using the derived optimal constellation design overlaps with the simulation results at any SNR in Case~II. We always use the optimal asymmetric constellation designs for these simulations, $\C_i = \{c_{0,i}, c_{1,i}\} = \{-\beta P_i,\; \alpha P_i\}$ for $i\in \{1,2\}$. To calculate the theoretical error probability, the decision boundaries are calculated by numerically solving for the roots of \eqref{xBoundExponential}. Then, these values are used to calculate the appropriate error expression, \eqref{Error1Solution} or \eqref{error3Solution}, based on the number of roots.

\begin{figure}[hbtp]
\centerline{\includegraphics[scale=0.43]{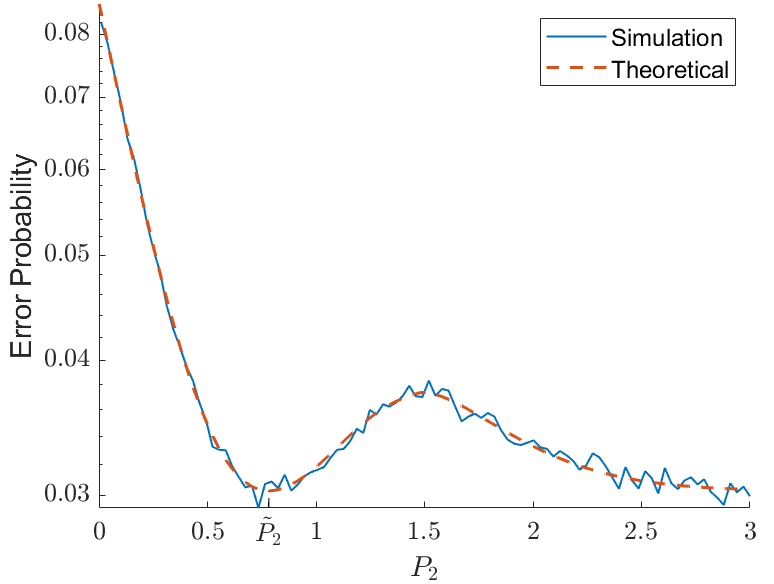}}
\caption{Theoretical and simulated error probability in Case~III ($p_1 = 0.45, \epsilon_1 = 0.01, \epsilon_2 = 0.05, P_1= 1, N_0=1$).}
\label{simulationCase3P2}
\end{figure}

The error probability as a function of $P_2$ and SNR are shown in Figs.~\ref{simulationCase3P2} and \ref{case2OtrhoCompare}, respectively. These plots show that the simulated and theoretical error performance overlaps very well, while also noting that the simulated minimum power allocation for $P_2$ coincides with the theoretical results.

\subsection{Simulated Comparison to Orthogonal Signaling}
\begin{figure}[hbtp]
\centerline{\includegraphics[scale=0.44]{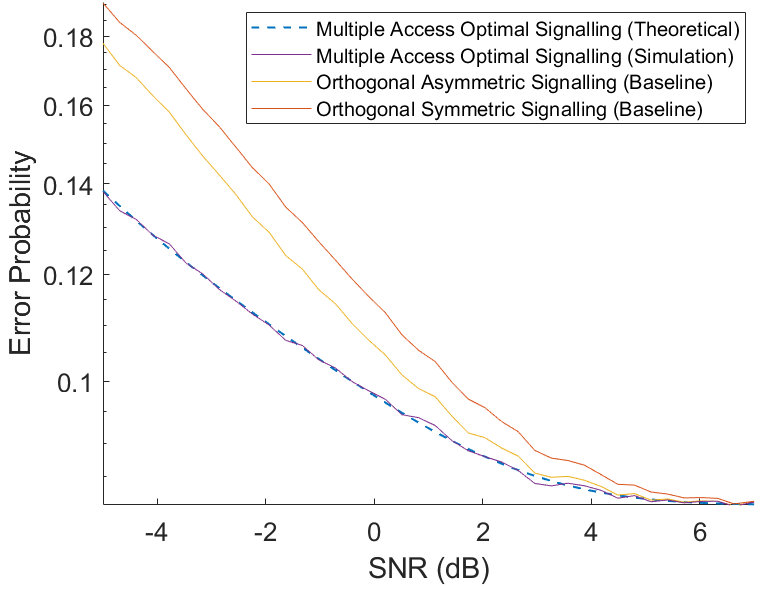}}
\caption{Error probability as a function of SNR in Case~II ($p_1 = 0.3, \epsilon_1 = 0.1, \epsilon_2 = 0.15, P_1^{\text{max}} = 1, P_2^{\text{max}} = 1$).}
\label{case2OtrhoCompare}
\end{figure}

\begin{figure}[hbtp]
\centerline{\includegraphics[scale=0.44]{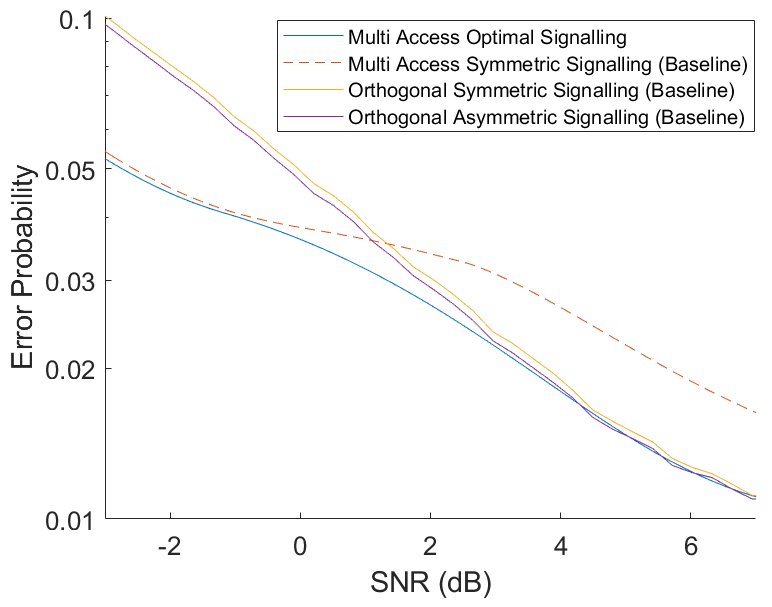}}
\caption{Error probability as a function of SNR in Case~III ($p_1 = 0.4, \epsilon_1 = 0.01, \epsilon_2 = 0.05, P_1^{\text{max}} = 1, P_2^{\text{max}} = 2$).}
\label{case3OtrhoCompare}
\end{figure}

In this section, we compare the error probability of the MAC signaling system derived in this paper to an alternative signaling method of using independent (orthogonal) channels for the sensors. To set up the orthogonal signaling, it is assumed that the sensor network would have access to two independent zero-mean Gaussian communication channels with variance~$\frac{N_0}{2}$. Note that we define the SNR in the orthogonal case to be the same as in \eqref{snrDef}. Even though there is more total noise when considering both orthogonal channels, this is a realistic comparison. If a system has access to two orthogonal channels with the same noise power, it can choose to only use one of the channels, which is exactly the equivalent MAC we are comparing to. We use two variations of orthogonal constellations as baseline comparisons. First, we use a simple symmetric binary phase-shift keying (BPSK) constellation design (i.e., $\C_i = \{c_{0,i}, c_{1,i}\} = \{-\sqrt{P_i^{\text{max}}},\; \sqrt{P_i^{\text{max}}}\}$ for $i\in \{1,2\}$). We also use the results of \cite{weng18} which give an optimal orthogonal constellation design to be asymmetric BPSK with $\C_i = \{c_{0,i}, c_{1,i}\} = \{-\beta\sqrt{P_i^{\text{max}}},\; \alpha\sqrt{P_i^{\text{max}}}\}$ for $i\in \{1,2\}$, with $\alpha$ and $\beta$ as defined in \eqref{alphaBetaDefs}. To detect the source, the receiver uses the MAP detection rule which is the two dimensional extension of \eqref{MAPDecodeExpression}. Since we have not analyzed the orthogonal channels case theoretically, we rely on simulation results to draw conclusions.  In Figs.~\ref{case2OtrhoCompare} and~\ref{case3OtrhoCompare}, the error probabilities are compared under two parameter sets, to show Cases II and III, respectively. Each data point is generated from 500,000 independent simulated source bits being sent through the channel. Fig.~\ref{case3OtrhoCompare}, also includes the error probabilities associated with using the maximum power symmetric constellation design over the MAC.

From these graphs we can see that in both Cases II and III, the derived optimal MAC constellation has better error performance than orthogonal signaling. However, in Case~III (Fig.~\ref{case3OtrhoCompare}), orthogonal signaling can perform better than the sub-optimal multiple access symmetric constellation design. These results
demonstrate that using a MAC optimally can have increased performance, while using less power and bandwidth. In Fig.~\ref{case2OtrhoCompare}, the maximum SNR gain of the derived optimal constellation compared to the next best option is about 2.4~dB. In Fig.~\ref{case3OtrhoCompare}, the maximum SNR gain is approximately 0.97~dB, occurring around 0.036 error probability.

\subsection{Analysis of Cases Based on Parameters $p_1, \epsilon_1$ and $\epsilon_2$}
\begin{figure}[htbp]
\centering
\begin{subfigure}{.5\linewidth}
  \centering
  \includegraphics[scale=0.215]{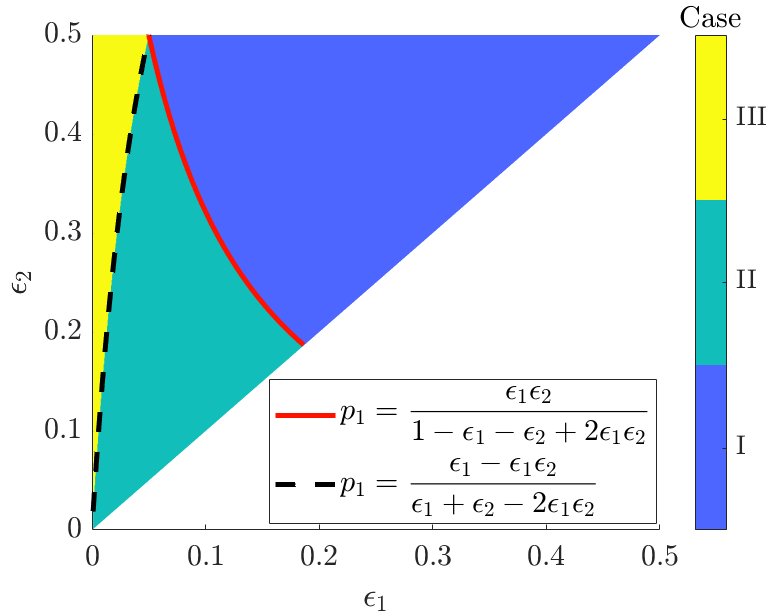}
  \caption{$p_1 = 0.05$}
  \label{caseTypeRegionsP0.05}
\end{subfigure}%
\begin{subfigure}{.5\linewidth}
  \centering
  \includegraphics[scale=0.215]{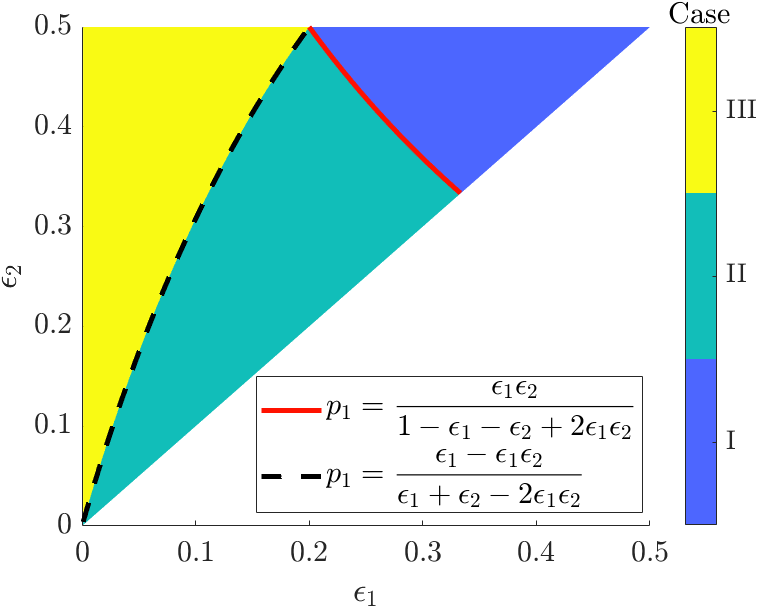}
  \caption{$p_1 = 0.2$}
  \label{caseTypeRegionsP0.2}
\end{subfigure}
\begin{subfigure}{.5\linewidth}
  \centering
  \includegraphics[scale=0.215]{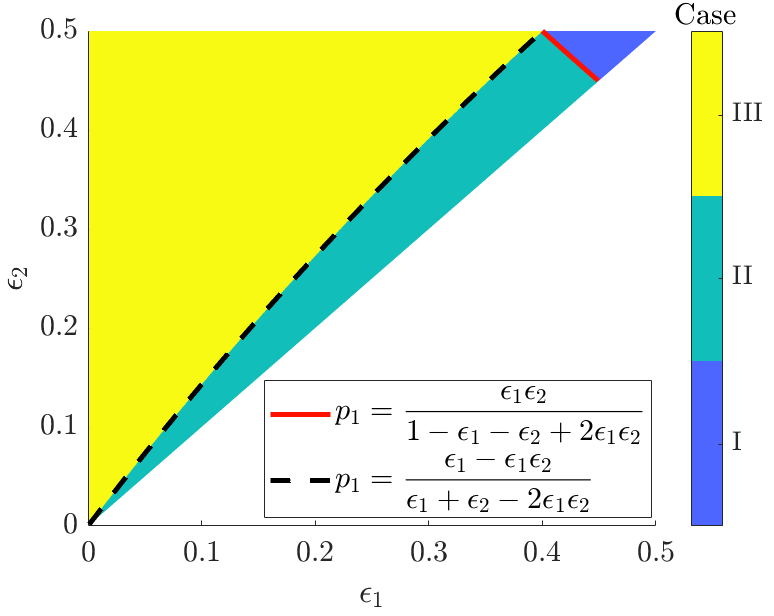}
  \caption{$p_1 = 0.4$}
  \label{caseTypeRegionsP0.4}
\end{subfigure}%
\begin{subfigure}{.5\linewidth}
  \centering
  \includegraphics[scale=0.215]{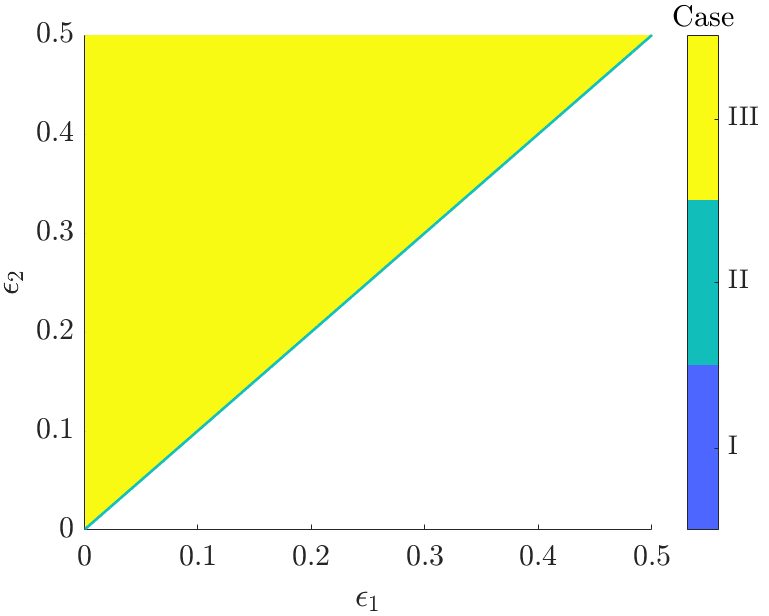}
  \caption{$p_1 = 0.5$}
  \label{caseTypeRegionsP0.5}
\end{subfigure}
\caption{Case type regions for different values of $p_1$.}
\label{caseTypeRegions}
\end{figure}

We analyze the behaviour of Cases~I-III as a function of the parameters $\epsilon_1$, $\epsilon_2$ and $p_1$. By fixing $p_1$, we illustrate the case type regions as a colour map of $\epsilon_1$ and $\epsilon_2$. Examples of these graphs are shown in Fig.~\ref{caseTypeRegions}.

We make the following observations from these diagrams. Case~I occurs at large $\epsilon_1$ and $\epsilon_2$ values, while Case~III is characterized by small $\epsilon_1 $ and large $\epsilon_2$. The boundaries between these regions are given exactly by the threshold equations given in Table~\ref{caseCharConditions}. As $p_1$ increases, the Case~I region becomes smaller, while Case~III becomes larger. Finally, at $p_1=0.5$, Case~I disappears entirely, and Case~II is equivalent to $\epsilon_1 =\epsilon_2$. This can intuitively be explained by noting that for any $p_1 < 0.5$, as $\epsilon_1, \epsilon_2 \rightarrow 0.5$, $X_1$ and $X_2$ become uniformly distributed and independent from the source $X$. This effectively removes the source information, making it useless to send over the channel (Case I). However, if $p_1 = 0.5$, $X_1$ and $X_2$ are uniformly distributed for any $\epsilon_1$ and $\epsilon_2$, so there is no statistical redundancy in the source (as it is unbiased) that can be lost when observed by the sensors. Hence it is always beneficial to send the signals, which explains why Case~I disappears at $p_1 = 0.5$.

\subsection{Error Performance vs. Power Allocation ($P_1$ and $P_2$)}
\begin{figure}[hbtp]
\centerline{\includegraphics[scale=0.43]{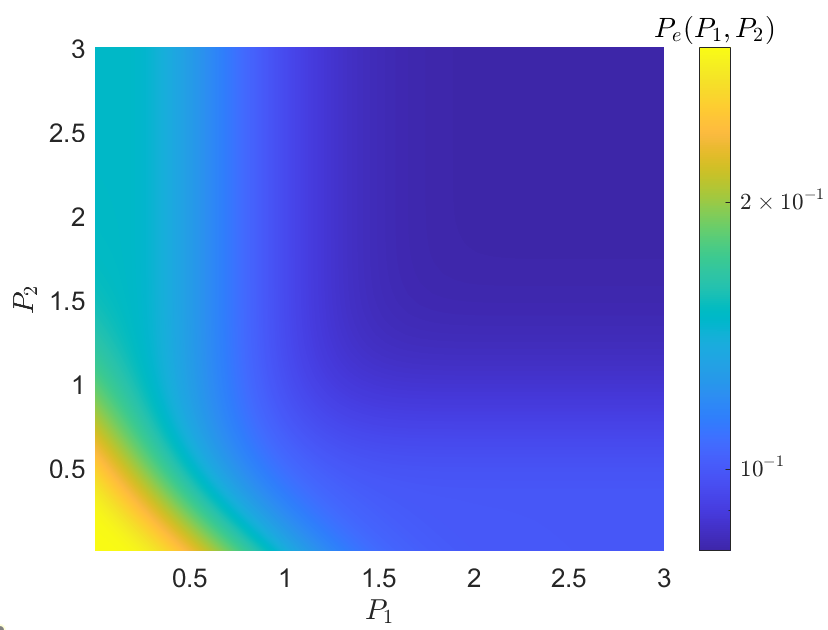}}
\caption{Error probability as a function of $P_1$ and $P_2$ in Case~II ($p_1 = 0.3, \epsilon_1 = 0.1, \epsilon_2 = 0.15, N_0 = 1$).}
\label{case2errorHeatMap}
\end{figure}

\begin{figure}[hbtp]
\centerline{\includegraphics[scale=0.43]{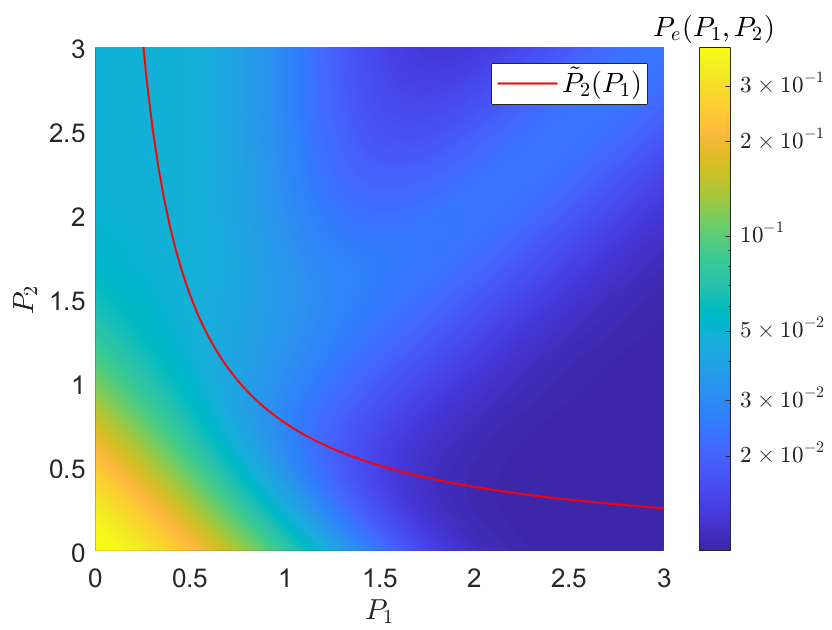}}
\caption{Error probability as a function of $P_1$ and $P_2$ in Case~III ($p_1 = 0.4, \epsilon_1 = 0.01, \epsilon_2 = 0.05, N_0 = 1$).}
\label{case3errorHeatMap}
\end{figure}

For the following examples, the constellations are parameterized by the optimal asymmetric design, $\C_i = \{c_{0,i}, c_{1,i}\} = \{-\beta P_i,\; \alpha P_i\}$, $i \in {1,2}$. Figs.~\ref{case2errorHeatMap} and \ref{case3errorHeatMap} show the error probability as a function of $P_1$ and $P_2$ in Cases II and III, respectively.

In Fig.~\ref{case2errorHeatMap}, increasing $P_1$ and $P_2$ always decreases the error probability, which reinforces the dervied optimal power allocation to use all available power. Fig.~\ref{case3errorHeatMap} illustrates the following properties of Case~III. First, we can see that for any fixed $P_1$ (vertical slice of the graph), the minimum occurs at $P_2 = \Tilde{P}_2(P_1)$, the red curve. Further, moving upward from $P_2 = 0$ to $P_2 = \Tilde{P}_2(P_1)$, we also see that the error probability decreases in $P_2$. Finally, we can see that moving rightward along the optimal power allocation curve, $\Tilde{P}_2(P_1)$, the optimal error probability decreases with $P_1$, which reinforces exactly the same optimal power allocation as proven. For example, if $P_1^\text{max} = P_2^\text{max} = 1$, then reading Fig. \ref{case3errorHeatMap} shows the optimal power allocations are $P_1^* = 1$ and $P_2^* = \Tilde{P}_2(1) \approx 0.76$. 

\subsection{Error Probability vs. Signal to Noise Ratio}

\begin{figure}[hbtp]
\centerline{\includegraphics[scale=0.4]{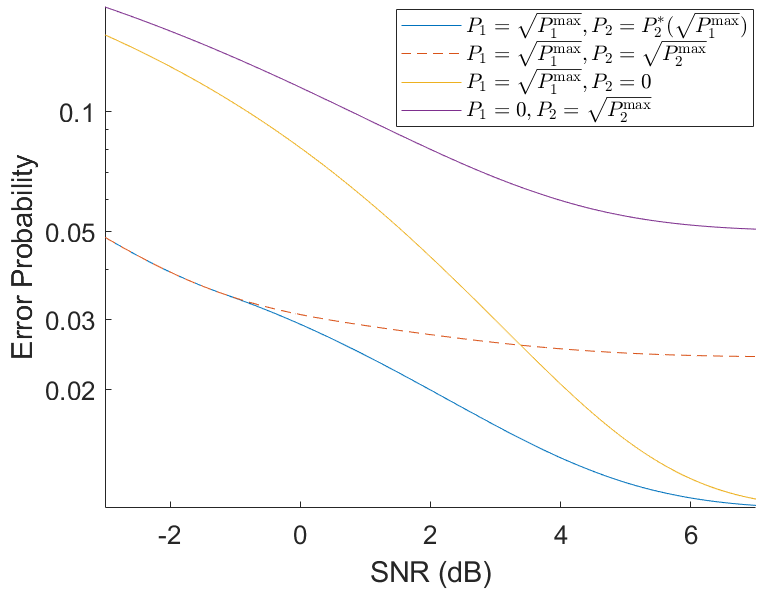}}
\caption{Error probability as a function of SNR in Case~III ($p_1 = 0.4, \epsilon_1 = 0.01, \epsilon_2 = 0.05, P_1^{\text{max}} = 1, P_2^{\text{max}} = 1$).}
\label{case3errorVarySNR}
\end{figure}

To demonstrate this system's SNR response, we vary $N_0$ to produce various SNR values as defined in \eqref{snrDef}. We also compare the derived optimal constellation design to other common power allocations. For the following example, the constellations are parameterized by the optimal asymmetric design, $\C_i = \{c_{0,i}, c_{1,i}\} = \{-\beta P_i,\; \alpha P_i\}$, $i \in {1,2}$. Fig.~\ref{case3errorVarySNR} shows the error probability of various constellation designs in Case~III as a function of $N_0$, expressed in terms of the SNR.

We make the following observations from this plot. First, at low SNR, the optimal and both max curves are identical. This is because for large enough values of $N_0$, $\sqrt{P_2^{\text{max}}} < \Tilde{P}_2$ as defined in \eqref{nonUniformP2TildeMax}, so $P_2^*(\sqrt{P_1^{\text{max}}}) = \sqrt{P_2^{\text{max}}}$. Next, at high SNR, the optimal and $P_1$ max curves become asymptotically equal. This is because as $N_0 \to 0$, $P_2^*(\sqrt{P_1^{\text{max}}}) \to 0$. At intermediate SNR (around 0-7 dB in this case), the optimal power allocation performs better than any of the alternatives. The largest SNR gain of using the derived optimal constellation is about 2.7~dB, occurring around 0.026 error probability.

\section{Conclusions and Future Work} \label{Conclusions}
In this paper, the optimal one dimensional constellation design for a two sensor binary network was established. After reducing the problem to a power allocation optimization problem (with the appropriate asymmetric constellation designs from Theorem~\ref{asymmetricParameter}), it was proved that there are distinct cases that arise based on the fixed parameters of the problem, which are $p_1$, $\epsilon_1$ and $\epsilon_2$. In some cases (Cases~I and~II), the results are intuitive and not unexpected, as the optimal power allocations are to use none or all of the available power. However, in Case~III, the most interesting and counter-intuitive result is that the optimal power allocation can be for Sensor~2 (with less correlation to the true data source) to use a portion, but not all of its available power. This is a significant result since Case~III is prevalent for many parameter sets of the problem setup. As shown in Fig.~\ref{caseTypeRegions}, Case~III becomes the dominant case as the binary source approaches a uniform distribution.
 
Although this result is theoretically significant, it is also important to consider the following details for any practical implications of this optimization. First, we note that in general, the decision boundaries may not be analytically solvable as they require solving for the roots of \eqref{xBoundExponential}. This means that it is most feasible to perform the calculations for a specific model of this problem ahead of time, and it would require complex computations to update the model while in use. Further, the optimal power allocation is sensitive to the parameters in the problem, so estimation errors for the parameters of a model could lead to worse performance than using a more robust design, such as using all available power. For this reason, the most practical implementation would be in a case where the optimal error has a large separation from the simpler constellation designs (around 0-5 dB in Fig.~\ref{case3errorVarySNR}). 

To expand upon these results, the following future directions can be considered. First, a natural extension is to generalize the problem to $N$ sensors ($N\ge 2$), each with their own correlation and power allotments. The analysis becomes increasingly complex for more than two sensors, so an approximation or bound on error performance could be investigated in some form of pairwise fashion using the results of this paper. Also, sensor network clustering problems such as those found in \cite{PEACH, radioCluster, kHop, LEACH-S} could be considered. The performance of a cluster could be approximated using the pairwise expressions derived in this paper. The balance between error probability and energy efficiency could be investigated as a function of cluster organization. Finally, we could consider a sensor network with $N$ sensors, but only two sensors send their data at any given time. The results of this paper could be used to decide which pairs of sensors should be selected, to stay within some power or error constraints.

\appendices
\section{Proof of Proposition~\ref{xInequalitiesCase2}}\label{ProofxInequalitiesCase2}
We prove \eqref{LIneqCase2} using the definition of $L$ from \eqref{LeqCase2}. Assume $P_1, P_2 > 0$, $\Bar{b} \neq 0$, then if $x$ satisfies
\begin{align*}
    x &\lesseqgtr \alpha P_1 - L(P_2)\\
    \implies ae^{\frac{2(\alpha + \beta)P_2x}{N_0}} + b &\lesseqgtr ae^{\frac{2(\alpha + \beta)P_2(\alpha P_1 - L(P_2))}{N_0}} + b,
\end{align*}
where
\begin{align*}
    &\quad\; ae^{\frac{2(\alpha + \beta)P_2(\alpha P_1 - L(P_2))}{N_0}} + b\\
    &= ae^{\frac{2\alpha(\alpha + \beta)P_1P_2 - 2(\alpha+\beta)P_2L(P_2)}{N_0}} + b\\
    &= ae^{\frac{2\alpha(\alpha + \beta)P_1P_2 + (\alpha^2-\beta^2)P_2^2}{N_0}-\ln\frac{\Bar{a}}{-\Bar{b}}} + b\\
    &=-\underbrace{\Bar{b}e^{\frac{-(\alpha P_1 - \beta P_2)^2}{N_0}}}_{=b} \underbrace{\frac{a}{\Bar{a}}e^{\frac{\alpha^2(P_1+P_2)^2}{N_0}}}_{=1}+ b\\
    &= -b +b = 0.
\end{align*}
The proof of \eqref{KIneqCase2} is omitted as it follows the same steps as above, using the definition of $K$ from \eqref{KeqCase2}. \qed
\section{Proof of Proposition~\ref{PropCase2P1XInfBound}} \label{P1PrimeIneqProof}
To prove \eqref{Case2P1XInfBound}, let $\Bar{P}_1$ and $P_1'$ be arbitrary real numbers such that $0<\Bar{P}_1<P_1'$. For any $P_1\in[\Bar{P}_1,P_1']$ and $P_2>0$, the root of \eqref{xBoundExponential}, $x$, satisfies $x - \alpha P_1 + L(P_2) > 0$ from \eqref{LBoundCase2} of Proposition~\ref{xBoundCase2}; therefore
\begin{equation} \label{infGeqZero}
    \inf_{P_1 \in [\Bar{P}_1,P_1']} x - \alpha P_1 + L(P_2) \geq 0.
\end{equation}
Next, we define the function
\begin{equation*}
    w_L(x,P_1) \triangleq ae^{\frac{2(\alpha+\beta)(P_1+P_2)x}{N_0}} + be^{\frac{2(\alpha+\beta)P_1x}{N_0}},
\end{equation*}
which is uniformly continuous in both $P_1$ over $[\Bar{P}_1,P_1']$ and $x$ over $[\alpha \Bar{P}_1 - L(P_2), \alpha P_1' - L(P_2)]$. By \eqref{LIneqCase2}, $w_L(x,P_1) = 0$ at all points $(x, P_1)$ on the line $x = \alpha P_1 - L(P_2)$. Thus, for any $d' > 0$, there exists $\delta > 0$ such that for any $P_1\in[\Bar{P}_1,P_1']$ and $x$ that satisfies
\begin{equation} \label{xCloseToBoundCondition}
    \alpha P_1 - L(P_2) < x < \alpha P_1 - L(P_2) + \delta,
\end{equation}
we have
\begin{equation*}
    w_L(x,P_1) < d'.
\end{equation*}

If \eqref{infGeqZero} holds with equality, then there exists $P_1 \in [\Bar{P}_1,P_1']$ with corresponding root of \eqref{xBoundExponential}, $x$, that satisfies \eqref{xCloseToBoundCondition}. In particular, for $d' = -\Bar{d}e^{-\frac{\beta^2(P_1'+P_2)^2}{N_0}}$, which is a positive constant with respect to $P_1$ and $x$, we obtain
\begin{align*}
    & ae^{\frac{2(\alpha+\beta)(P_1+P_2)x}{N_0}} + be^{\frac{2(\alpha+\beta)P_1x}{N_0}} < -\Bar{d}e^{-\frac{\beta^2(P_1'+P_2)^2}{N_0}} < -d \\
    & \implies ae^{\frac{2(\alpha+\beta)(P_1+P_2)x}{N_0}} + be^{\frac{2(\alpha+\beta)P_1x}{N_0}} + d < 0 \\
    &\implies w(x) < 0,
\end{align*}
where the last inequality holds because $c<0$ in the expression of $w(x)$ in~\eqref{xBoundExponential} for Case~II. This contradicts $x$ being the root of \eqref{xBoundExponential}, completing the proof. The proof of \eqref{Case2P2XInfBound} is omitted as it follows the same steps as above. \qed 
\section{Proof of Proposition~\ref{FGminimization}} \label{ProofFGMinimization}
Let $P_1,P_2 > 0$. The following are expressions for the derivatives:
\begin{align*}
\frac{dg}{dx} &= \frac{1}{\sigma\sqrt{2\pi}}e^{\frac{-x^2}{N_0}}\bigg(ae^{\frac{2\alpha(P_1+P_2)x}{N_0}} + ce^{\frac{2(-\beta P_1 + \alpha P_2)x}{N_0}}\bigg) \\ 
\frac{dh}{dx} &= \frac{1}{\sigma\sqrt{2\pi}}e^{\frac{-x^2}{N_0}}\bigg(be^{\frac{2(\alpha P_1 - \beta P_2)x}{N_0}} + de^{\frac{-2\beta(P_1+P_2)x}{N_0}}\bigg).
\end{align*}
    Next, we apply the results of Proposition~\ref{xInequalitiesCase3} to show 
    \begin{align*}
        x &\lessgtr \alpha P_2 - K_\alpha(P_1) \implies 0 \;\reflectbox{$\lessgtr$}\; ae^{\frac{2(\alpha+\beta)P_1x}{N_0}} + c \\
        \implies 0 &\;\reflectbox{$\lessgtr$}\; ae^{\frac{2\alpha(P_1+P_2)x}{N_0}} + ce^{\frac{2(-\beta P_1 + \alpha P_2)x}{N_0}} \\
        \implies 0 &\;\reflectbox{$\lessgtr$}\;  \frac{dg}{dx},
    \end{align*}
    and
    \begin{align*}
        x &\lessgtr -\beta P_2 + K_\beta(P_1) \implies 0 \;\reflectbox{$\lessgtr$}\; be^{\frac{2(\alpha+\beta)P_1x}{N_0}} + d \\
        \implies 0 &\;\reflectbox{$\lessgtr$}\; be^{\frac{2(\alpha P_1 - \beta P_2)x}{N_0}} + de^{\frac{-2\beta(P_1+P_2)x}{N_0}} \\
        \implies 0 &\;\reflectbox{$\lessgtr$}\; \frac{dh}{dx}.
    \end{align*} \vspace{-20pt}\qed
\section{Proof of Theorem~\ref{Case3DecreaseP1}} \label{ProofCase3DecreaseP1}
Let $P_1 > 0$. We can express the optimal power allocation for $P_2$ as the following function of $P_1$:
    \begin{equation*}
    P_2^*(P_1) = 
    \begin{cases} 
      \sqrt{P_2^{\text{max}}} & P_1 <  P_1^{thresh} \\
      \Tilde{P}_2(P_1) & P_1 \geq P_1^{thresh}
    \end{cases}
    \end{equation*}
    where 
    \begin{equation*}
        P_1^{thresh} \triangleq \frac{N_0}{2(\alpha+\beta)^2\sqrt{P_2^{\text{max}}}}\ln\frac{\Bar{a}\Bar{d}}{\Bar{b}\Bar{c}}.
    \end{equation*}

    We analyze this in two cases. First, assume $P_1 < P_1^{thresh}$. Since $P_2^*(P_1) = \sqrt{P_2^{\text{max}}} < \Tilde{P}_2(P_1)$, Proposition~\ref{case3OneSolP2} implies that there is a unique root, $x$, to \eqref{xBoundExponential} corresponding to $P_1$ and $P_2$.  Hence, by the same reasoning as in Theorem~\ref{Case2DecreaseP1P2} it is sufficient to show $P_{e,x}^{\text{\tiny UB}}$ is decreasing in $P_1$. Unlike the previous proof, we can see immediately from the expression given in \eqref{PUBderivativeP1} that this derivative is negative for all $P_1$ because $\Bar{b}>0$ in Case~III. Next, for $P_1 \geq P_1^{thresh}$ we have
\begin{equation*}
    P_2^*(P_1) =  \frac{N_0}{2(\alpha+\beta)^2P_1}\ln\frac{\Bar{a}\Bar{d}}{\Bar{b}\Bar{c}}.
\end{equation*}
Let $x^*$ be the root to \eqref{xBoundExponential} for $P_1$ and $P_2^*(P_1)$. \eqref{xBoundsP2Case3Eqn} implies $x^* = \alpha P_2^*(P_1) - K_\alpha (P_1) = -\beta P_2^*(P_1) + K_\beta (P_1)$. Substituting these relationships into the expression for the error probability given in \eqref{Error1Solution} yields
\begin{align*}
    P_e\big(P_1, P_2^*(P_1)\big) 
    &= \Bar{a}Q\Bigg(\frac{1}{\sigma}\bigg(\alpha P_1 + P_a - \frac{\alpha - \beta}{2}P_1\bigg)\Bigg)\\
    &\quad+ \Bar{c}Q\Bigg(\frac{1}{\sigma}\bigg(-\beta P_1 + P_a - \frac{\alpha - \beta}{2}P_1\bigg)\Bigg)\\
    &\quad+\Bar{b}Q\Bigg(\frac{1}{\sigma}\bigg(\alpha P_1 - P_b - \frac{\alpha - \beta}{2}P_1\bigg)\Bigg) \\
    &\quad+ \Bar{d}Q\Bigg(\frac{1}{\sigma}\bigg(-\beta P_1 - P_b - \frac{\alpha - \beta}{2}P_1\bigg)\Bigg)\\
    &\quad+ \hspace{-10pt}\sum_{(l,m)\in\{0,1\}^2}\hspace{-10pt}p_0p_{lm|0},
\end{align*}
where we define
\begin{align*}
    P_a &\triangleq \frac{N_0}{2(\alpha+\beta)P_1}\ln\frac{\Bar{a}}{-\Bar{c}},\quad  P_b &\triangleq \frac{N_0}{2(\alpha+\beta)P_1}\ln\frac{-\Bar{d}}{\Bar{b}},
\end{align*}
noting that 
\begin{equation*}
    \frac{dP_a}{dP_1} = -\frac{P_a}{P_1}, \qquad \frac{dP_b}{dP_1} = -\frac{P_b}{P_1}.
\end{equation*}
Then derivative analysis on the error probability yields
\begin{align*}
    \frac{dP_e}{dP_1} &= -\frac{1}{\sigma\sqrt{2\pi}}\Bigg(\bigg(\frac{\alpha+\beta}{2}-\frac{P_a}{P_1}\bigg)\Bar{a}e^{\frac{-(\frac{\alpha+\beta}{2}P_1 + P_a)^2}{N_0}} \\
    &\qquad\qquad+ \bigg(-\frac{\alpha+\beta}{2}-\frac{P_a}{P_1}\bigg)\Bar{c}e^{\frac{-(-\frac{\alpha+\beta}{2}P_1 + P_a)^2}{N_0}} \\
    &\qquad\qquad + \bigg(\frac{\alpha+\beta}{2}+\frac{P_b}{P_1}\bigg)\Bar{b}e^{\frac{-(\frac{\alpha+\beta}{2}P_1 - P_b)^2}{N_0}} \\
    &\qquad\qquad+\bigg(-\frac{\alpha+\beta}{2}+\frac{P_b}{P_1}\bigg)\Bar{d}e^{\frac{-(\frac{\alpha+\beta}{2}P_1 + P_b)^2}{N_0}}\Bigg) \\
    &= -\frac{e^{\frac{-(\frac{\alpha+\beta}{2}P_1)^2}{N_0}}}{\sigma\sqrt{2\pi}}\Bigg(e^{\frac{-P_a^2}{N_0}}\bigg(\Big(\frac{\alpha+\beta}{2}-\frac{P_a}{P_1}\Big)\Bar{a}e^{\frac{-(\alpha + \beta)P_1P_a}{N_0}} \\
    &\qquad\qquad+ \Big(-\frac{\alpha+\beta}{2}-\frac{P_a}{P_1}\Big)\Bar{c}e^{\frac{(\alpha + \beta)P_1P_a}{N_0}}\bigg) \\
    &\qquad\qquad + e^{\frac{-P_b^2}{N_0}}\bigg(\Big(\frac{\alpha+\beta}{2}+\frac{P_b}{P_1}\Big)\Bar{b}e^{\frac{(\alpha + \beta)P_1P_b}{N_0}} \\
    &\qquad\qquad+ \Big(-\frac{\alpha+\beta}{2}+\frac{P_b}{P_1}\Big)\Bar{d}e^{\frac{-(\alpha + \beta)P_1P_b}{N_0}}\bigg)\Bigg) \\
    &= -\frac{e^{\frac{-(\frac{\alpha+\beta}{2}P_1)^2}{N_0}}}{\sigma\sqrt{2\pi}}\Bigg(e^{\frac{-P_a^2}{N_0}}\bigg(\Big(\frac{\alpha+\beta}{2}-\frac{P_a}{P_1}\Big)\sqrt{-\Bar{a}\Bar{c}} \\
    &\qquad\qquad- \Big(-\frac{\alpha+\beta}{2}-\frac{P_a}{P_1}\Big)\sqrt{-\Bar{a}\Bar{c}}\big) \\
    &\qquad\qquad + e^{\frac{-P_b^2}{N_0}}\bigg(\Big(\frac{\alpha+\beta}{2}+\frac{P_b}{P_1}\Big) \sqrt{-\Bar{b}\Bar{d}} \\
    &\qquad\qquad- \Big(-\frac{\alpha+\beta}{2}+\frac{P_b}{P_1}\Big)\sqrt{-\Bar{b}\Bar{d}}\bigg)\Bigg) \\
    &= -\frac{\alpha + \beta}{\sigma\sqrt{2\pi}}e^{\frac{-(\frac{\alpha+\beta}{2}P_1)^2}{N_0}}\bigg(e^{\frac{-P_a^2}{N_0}}\sqrt{-\Bar{a}\Bar{c}} + e^{\frac{-P_b^2}{N_0}}\sqrt{-\Bar{b}\Bar{d}}\bigg)\\
    &< 0.
\end{align*}
\vspace{-20pt}
\qed

\balance
\bibliographystyle{IEEEtran}
\bibliography{references}
\end{document}